# The Radio Spectral Energy Distribution and Star Formation Calibration in MIGHTEE-COSMOS Highly Star-Forming Galaxies at $1.5 < z < 3.5$

Fatemeh Tabatabaei,[1,2,3] Maryam Khademi,[1,*] Matt J. Jarvis,[4,5] Russ Taylor,[5] Imogen H. Whittam,[4,5] Fangxia An,[6,5] Reihaneh Javadi,[1] Eric J. Murphy,[7] and Mattia Vaccari[8,9,10]

[1]*School of Astronomy, Institute for Research in Fundamental Sciences (IPM), P.O. Box 1956836613, Tehran, Iran*
[2]*Max-Planck-Institiut für Astronomy, Königstul 17, D-69117, Heidelberg, Germany*
[3]*Max-Planck Institut für Radioastronomie, Auf dem Hügel 69, D-53121 Bonn, Germany*
[4]*Sub-Department of Astrophysics, University of Oxford, Denys Wilkinson Building, Keble Road, Oxford, OX1 3RH, UK*
[5]*Department of Physics and Astronomy, University of the Western Cape, Robert Sobukwe Road, 7535 Bellville, Cape Town, South Africa*
[6]*Purple Mountain Observatory, Chinese Academy of Sciences, 10 Yuan Hua Road, Nanjing, Jiangsu 210023, People's Republic of China*
[7]*National Radio Astronomy Observatory, 520 Edgemont Road, Charlottesville, VA 22903, USA*
[8]*Inter-University Institute for Data Intensive Astronomy, Department of Astronomy, University of Cape Town, 7701 Rondebosch, Cape Town, South Africa*
[9]*Inter-University Institute for Data Intensive Astronomy, Department of Physics and Astronomy, University of the Western Cape, 7535 Bellville, Cape Town, South Africa*
[10]*INAF - Istituto di Radioastronomia, via Gobetti 101, 40129 Bologna, Italy*

## ABSTRACT

Studying the radio spectral energy distribution (SED) of distant galaxies is essential for understanding their assembly and evolution over cosmic time. We present rest frame radio SEDs of a sample of 160 starburst galaxies at $1.5 < z < 3.5$ in the COSMOS field as part of the MeerKAT International GHz Tiered Extragalactic Exploration (MIGHTEE) project. MeerKAT observations combined with archival VLA and GMRT data allow us to determine the integrated mid-radio ($\nu$=1-10 GHz) continuum (MRC) luminosity and magnetic field strength. A Bayesian method is used to model the SEDs and to separate the free-free and synchrotron emission. We also calibrate the star formation rate (SFR) in radio both directly through SED analysis and indirectly through the infrared–radio correlation (IRRC). With a mean value of $\alpha_{nt} \simeq 0.7$, the synchrotron spectral index flattens with both redshift and specific star formation rate (sSFR), indicating that cosmic rays are more energetic in the early universe due to higher star formation activity. The magnetic field strength increases with redshift, $B \propto (1+z)^{(0.7\pm 0.1)}$ and SFR as $B \propto \mathrm{SFR}^{0.3}$, suggesting a small-scale dynamo acting as its main amplification mechanism. Taking into account the evolution of the SEDs, the IRRC is redshift-invariant and it does not change with stellar mass at $1.5 < z < 3.5$ although the correlation deviates from linearity. Similarly, we show that the SFR traced using the integrated MRC luminosity is redshift invariant.

## 1. INTRODUCTION

The radio continuum (RC) emission from star forming galaxies originates mainly at different phases of massive star formation over cosmic time (e.g., Condon 1992). However, the RC emission as a dust-unbiased tracer of star formation rate (SFR) was first inferred through its globally tight correlation with the infrared (IR) emission (see e.g., Condon et al. 2002; Garrett 2002; Appleton et al. 2004; Sajina et al. 2008; Sargent et al. 2010; Leslie et al. 2020; Garn et al. 2009, and references therein). Calibrating the SFR using the IR–RC correlation (IRRC), however, is complicated by its observed change with redshift (e.g., Basu et al. 2015; Delhaize et al. 2017; Magnelli et al. 2015) which can depend on galaxy mass and type (e.g., Delvecchio et al. 2017, 2021a; Read et al. 2018; Molnár et al. 2021; De Zotti et al. 2024). Studies in nearby and local star forming galaxies show that the correlation can be non-linear (e.g., Niklas & Beck 1997; Dumas et al. 2011; Heesen et al. 2014; Matthews et al. 2021) and the correlation can change with the ISM conditions and spatial scale (e.g., Tabatabaei et al. 2013a,b). There are also studies questioning the direct connection of the IRRC, to star formation and instead suggest a conspiracy of several factors (e.g., Bell 2003a; Lacki et al. 2010). All these show the importance of 1) a better understanding of the

Corresponding author: Maryam Khademi
* Email: maryam.khademi@ipm.ir

properties of the RC radiation and its possible variation over cosmic time and 2) exploring direct and physically motivated methods of the SFR calibration using the RC emission. Such methods have been studied in nearby galaxies (Murphy et al. 2011; Tabatabaei et al. 2017) but not well explored at high redshifts.

Deep multi-frequency radio surveys of galaxies enable the analysis of their SEDs and the study of the radiation laws and physical processes taking place over cosmic times. In local star-forming galaxies, the RC emission emerges from two main radiation mechanisms, the thermal free-free emission (mainly from HII regions) and the non-thermal synchrotron radiation of cosmic rays spiraling along the interstellar magnetic field (which also originates mainly in star-forming regions Condon 1992; Tabatabaei et al. 2022). A detailed study of the radio SEDs of the KINGFISH (Key Insights on Nearby Galaxies: a Far-IR Survey with Herschel, Kennicutt et al. 2011) sample of normal star-forming galaxies showed that the nonthermal emission follows a power-law relation with a spectral index of $\alpha_{nt} = 0.97 \pm 0.16$ and the thermal fraction (thermal-to-observed flux ratio) is responsible for about 23% of the RC luminosity integrated over the frequency range 1 to 10 GHz (the so-called mid-radio luminosity, MRC) on average (Tabatabaei et al. 2017). The nonthermal SEDs become flatter in galaxies with higher star formation surface density ($\Sigma_{\rm SFR}$) likely due to a younger and more energetic population of cosmic ray electrons (Tabatabaei et al. 2017). In starburst galaxies, the thermal fraction can be relatively lower than in normal star-forming galaxies due to an efficient amplification of magnetic fields and acceleration of cosmic ray electrons (Tabatabaei et al. 2017; Lacki & Beck 2013; Condon 1992; Krumholz et al. 2020). These indicate that the intrinsic shape of the radio SEDs does not remain unchanged with redshift due to the evolution of the SFR and the ISM conditions (Ghasemi-Nodehi et al. 2022). While such variations can have an important consequence on the IRRC, and SFR calibrations, they are often neglected in high-z studies or not properly addressed (Tisanić et al. 2019; An et al. 2021).

Exploring the direct use of the RC emission to calibrate the SFR in nearby galaxies, Tabatabaei et al. (2017) showed that the integrated MRC luminosity provides a more robust SFR tracer than the single-band radio luminosities often used. Deep field multi-band RC observations are vital for similar studies in the distant universe. Furthermore, comparing the SEDs in the radio and IR provides key insights into the nature of the emission and general factors balancing their energy. As we approach the epoch of the multi-frequency radio surveys with the SKA, ngVLA, and pathfinders, we can address these properly by measuring radio SEDs and comparing them with available IR SEDs of various astrophysical objects. We will also not be limited by a fixed non-thermal spectral index which cannot explain either the resolved spectra of galaxies (e.g., Tabatabaei et al. 2007, 2013c) or the integrated spectra (e.g., Duric et al. 1988; Marvil et al. 2015).

Detailed SED analyses can help dissect the thermal and nonthermal processes that control the ISM energy balance and structure formation. Observations in nearby galaxies show that the non-thermal pressures inserted by the magnetic fields and relativistic particles can be higher than the thermal pressures (Beck 2007; Hassani et al. 2022) and can play a role in decelerating the SFR (Tabatabaei et al. 2018, 2022). However, information about the importance of these processes in the early universe and at high redshifts is still lacking.

Thanks to its rich auxiliary data available, the Cosmic Evolution Survey (COSMOS)[1] allows multi-wavelength studies of galaxies at high redshifts. The MeerKAT International GHz Tiered Extragalactic Exploration (MIGHTEE, Jarvis et al. (2016)) is a large key science project to study the properties and evolution of galaxies over cosmic time by creating deep GHz radio continuum, spectral line, and polarisation observations. Combining the MIGHTEE Early Science data at 1.3 GHz with other radio surveys, An et al. (2021) presented observed radio spectral indices for a sample of star-forming galaxies (SFGs) in the COSMOS field by using their flux ratios. In this paper, we present a more detailed SED analysis for a sub-sample of those SFGs located at high redshifts ($1.5 < z < 3.5$) in order to disentangle their thermal and nonthermal emission properties and obtain their rest-frame MRC. This analysis also allows us to estimate the equipartition magnetic field strength, revisit the evolution of the IRRC, and obtain more direct SFR calibration relations.

This paper is organized as follows. After describing the data and the sample in Sect. 2, we explain the SED modeling, the fitting algorithm, and the results in Sect. 3. The radio SEDs are then integrated to obtain the MRC luminosity (Sect. 4) and magnetic field strength (Sect. 5). Cosmic evolution of the radio SEDs and its effect on the IRRC as well as the SFR calibrations are presented and discussed in Sect. 6 and summarized in Sect. 7.

## 2. DATA AND SAMPLE SELECTION

---

[1] https://cosmos.astro.caltech.edu/

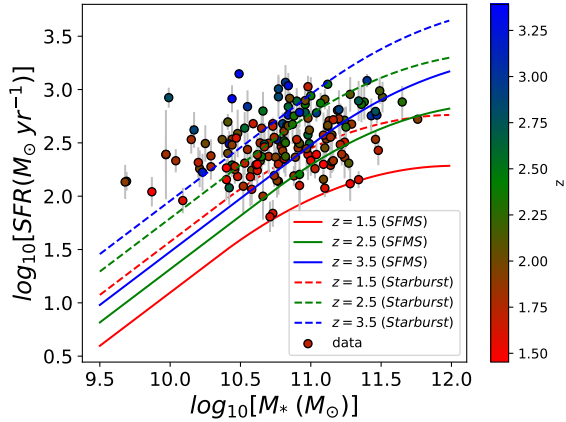

Figure 1: Star formation rate (SFR) against stellar mass ($M_\star$) of a sample of 160 SFGs in the MIGHTEE-COSMOS field located at $1.5 < z < 3.5$. Galaxies are redshift color-coded. Also shown are the log(SFR)-log($M_\star$) scaling relations of the main sequence star forming galaxies (*solid curves*, Schreiber et al. 2015) at selected redshifts and their $3\times$ shifts (*dashed curves*) as a criteria above which galaxies of the same redshift (color) are counted as starbursts (see e.g., Cooke et al. 2023, and references therein).

Table A1 lists the multi-wavelength data of a subsample of superdeblended catalog of the MIGHTEE-COSMOS SFGs (An et al. 2021) at $z > 1.5$ used in this study. A summary of related observations and the selection of the sample are described as follows.

### 2.1. MIGHTEE MeerKAT Observations

MIGHTEE will cover 20 square degrees in four well-studied extragalactic fields, such as the COSMOS field, to $\mu$Jy sensitivity at 1.3 GHz (Heywood et al. 2022). As part of the MIGHTEE project, the MeerKAT telescope, a state-of-the-art 64-antenna radio array located in the Northern Cape of South Africa, observed the inner 1.6 deg$^2$ area of the COSMOS field (RA= $10^h\,00^m\,28.6^s$, DEC=+$2°\,12'\,21''$) with a single deep pointing (17.45 h on-source integration) in L-band centered at 1.3 GHz. This paper uses the Early Science Continuum data of MIGHTEE covering $\sim 1.6$ deg$^2$ of the COSMOS field released by Heywood et al. (2022). After reduction and calibration, the radio continuum image created using Robust=0 weighting resulted in an angular resolution of $8.6''$ reaches a thermal noise of $1.7\,\mu$Jy beam$^{-1}$. However, due to classical confusion, i.e., a contribution from the numerous faint and unresolved astronomical sources, the nominal $1\sigma$ rms of the map increases to $5.5\,\mu$Jy beam$^{-1}$ (Heywood et al. 2022). As explained by (Heywood et al. 2022), a total of 9896 radio sources were found above $5\sigma$ in the primary beam corrected map.

### 2.2. VLA Data

The VLA 1.4 GHz data are a combination of the VLA-COSMOS Large Project covering the entire 2 deg$^2$ with a mean rms noise of $15\,\mu$Jy beam$^{-1}$ (Schinnerer et al. 2007) and the VLA-COSMOS Deep Project (Schinnerer et al. 2010) covering the central $50' \times 50'$ in the COSMOS field with a maximum sensitivity of $12\,\mu$Jy per $1''.5 \times 1''.4$ beam. From this combined imaging data, 2856 radio sources were detected with a signal-to-noise ratio (SNR) $> 5$ (peak flux densities, see Schinnerer et al. 2010, for details). At 3 GHz, we used the Karl G. Jansky VLA-Cosmic Evolution Survey 3GHz Large Project which mapped the entire 2 deg$^2$ COSMOS field with a median rms of $2.3\,\mu$Jy per $0.75''$ beam (Smolčić et al. 2017a). In total, 10830 radio sources were detected at $> 5\sigma$ as described by Smolčić et al. (2017a).

### 2.3. GMRT Data

The Giant Meter-wave Radio Telescope (GMRT) observations covering the entire 2 deg$^2$ COSMOS field with a channel width of 125 kHz, and a total bandwidth of 32 MHz, were carried out using 30 antennas, in which the longest baseline is 25 km (Tisanić et al. 2019). The GMRT observations at 610 MHz and 325 MHz have been taken from Tisanić et al. (2019), with source flux densities down to $5\sigma$ rms detection level. Median rms values of 97 and 39 $\mu$Jy beam$^{-1}$ were measured for observations at 325 MHz and 610 MHz, respectively (Smolčić et al. 2014; Tisanić et al. 2019). The 610 MHz observations were carried out at a central frequency of 608 MHz with a bandwidth of 32 MHz, covering the entire 2 deg$^2$ COSMOS field whose data have been imaged at a resolution of $5.6'' \times 3.9''$. At 325 MHz, observations of a single pointing of the entire COSMOS field were performed with a bandwidth of 32 MHz whose data have been imaged at a resolution of $10.8'' \times 9.5''$.

### 2.4. Other Data

In addition to the radio data, we used the data of the total infrared luminosity ($L_{\rm TIR}$) and SFR of galaxies in the COSMOS field (Jin et al. 2018; Smolčić et al. 2017b). As explained by An et al. (2021), the authors presented a super-deblended IR to (sub)millimeter photometry from Spitzer (Floc'h et al. 2009), Herschel (Oliver et al. 2010; Lutz et al. 2011; Béthermin et al. 2012), SCUBA2 (Cowie et al. 2017; Geach et al. 2016), AzTEC (Aretxaga et al. 2011), and MAMBO (Bertoldi et al. 2007) images in the COSMOS field. These data allowed Jin et al. (2018) to construct the IR to millimeter SEDs of galaxies. They obtained $L_{\rm TIR}$ by integrating the SEDs over the 8-1000 $\mu$m frequency range. Utilizing the superdeblended photometry, Jin et al. (2018)

conducted SED fittings from FIR to millimeter wavelengths for these sources. They estimated their star formation rates (SFR) by integrating the IR luminosities ($L_{\rm IR}$) across the 8-1000 $\mu$m range, derived from the best-fitting SEDs, by assuming a Chabrier IMF (Chabrier 2003).

The COSMOS field has been extensively studied in the optical, UV, and Infrared wavelengths with the Suprime-Cam/Subaru (Taniguchi et al. 2007, 2015), the MegaCam/CFHT ($u^\star$-band data from Canada–France–Hawaii Telescope), GALEX (Zamojski et al. 2007), VIRCAM/VISTA (UltraVISTA-DR2)[2], HSC/Subaru (Miyazaki et al. 2012), WIRCam/CFHT (McCracken et al. 2009) and IRAC/Spitzer (SPLASH Hsieh et al. 2012) surveys. The photometric redshifts were taken from the COSMOS2020 catalogue (Weaver et al. 2022) for most of our sample. We also used the older COSMOS2015 catalog (Laigle et al. 2016) for few galaxies with no photometric redshift information in the COSMOS2020 catalog. These catalogs also include information on stellar mass estimated by fitting the Mid-IR to near-UV SEDs of the galaxies assuming the Chabrier IMF. The accuracy in photometric redshifts ($z_{\rm p}$) was estimated by comparing the photometric redshifts with robust spectroscopic measurements ($z_{\rm s}$) which is defined as $\sigma_{\Delta z/(1+z)} = 1.48 \times$ median($|z_{\rm p} - z_{\rm s}|/(1 + z_{\rm s})$) resulting in an accuracy of $\sigma_{\Delta z/(1+z)} = 0.007$ to $0.069$ depending on the spectroscopic survey used as the reference (Laigle et al. 2016; Weaver et al. 2022, see Table. 5).

### 2.5. *Sample Selection*

The frequency range of the RC data available (0.3 < $\nu$ < 3 GHz) allows us to study the rest-frame mid-radio SEDs for galaxies at $z > 1.5$. Hence, we construct our sample using the sample of SFGs defined by An et al. (2021)[3] with $z > 1.5$. Moreover, for a robust SED analysis, our sample is limited to only detections at all RC frequencies (Sect. 3.1). A brief description of An et al.'s sample selection is presented below, and we refer the reader to their paper for further details.

An et al. (2021) matched the photometry of the MeerKAT survey at 1.3 GHz with the photometry from the VLA at two frequencies of 1.4 GHz and 3 GHz and those of GMRT at 0.3 GHz and 0.6 GHz by using the COSMOS Super-deblended Photometry Catalog which is a highly complete prior sample of 194428 galaxies presented by Jin et al. (2018). The SFGs were first selected among the whole 4509 VLA 3 GHz sources with SNR > 5 within the area of the MeerKAT primary beam by removing sources previously classified as AGNs or red galaxies. This was performed by cross-matching the VLA 3 GHz radio sources with optical/NIR counterparts of X-ray sources from Marchesi et al. (2016) inside a 1-arcsecond radius. This revealed that 566 out of 4509 radio sources have Chandra X-ray detection, with 519 identified as AGNs based on their rest-frame X-ray luminosity ($L_X > 10^{42}$ $erg$ $s^{-1}$) from Marchesi et al. (2016). Additionally, they cross-correlated these 3 GHz radio sources with optical/NIR counterparts of 2012 X-ray sources from the XMM–Newton X-ray survey (Cappelluti et al. 2009; Brusa et al. 2010), uncovering six more X-ray detected AGNs.

The selection of AGNs using X-ray data often overlooks low-luminosity AGNs at high redshifts because they generally lack X-ray emissions related to accretion (Marchesi et al. 2016). To address this, An et al. (2021) also used the MIR color-color selection criteria from Donley et al. (2012) for sources with redshifts ($z \leq 3$). The superdeblended catalog, which includes all VLA 3 GHz sources with a signal-to-noise ratio (SNR) greater than 5 (Jin et al. 2018), provides matched photometry. From this catalog, An et al. (2021) identified 2760 sources in their sample with SNR greater than 3 in all four IRAC bands. Out of these, 240 sources meet the MIR AGN selection criteria (Donley et al. 2012).

Jin et al. (2018) identified 726 out of 4509 radio-excess sources in the superdeblended catalog as radio-loud AGNs. The criteria for defining radio excess are determined as $(S_{obs} - S_{SED})/\sqrt{(\sigma_{obs}^2 + \sigma_{SED}^2)} > 3$ and $S_{obs} > 2 \times S_{SED}$ (Liu et al. 2018), in which $S_{obs}$ is the observed radio flux density and $S_{SED}$ is the radio flux density obtained by the best-fitting FIR to millimeter SEDs.

For the 566 and 311 VLA 3 GHz radio sources cross-matched with the optical/NIR counterparts of X-ray sources from Marchesi et al. (2016) and Brusa et al. (2010), respectively, 160 are identified as AGNs based on their best-fitting optical to NIR SED in Marchesi et al. (2016). Additionally, 392 and 183 sources are spectroscopically identified as AGNs in Marchesi et al. (2016) and Brusa et al. (2010), respectively. Furthermore, 3958 out of 4509 radio sources are included in the 5$\sigma$ 3 GHz catalog presented by Smolčić et al. (2017b). Among these, 501, 271, 578, and 793 are classified as X-ray, MIR, SED, and radio excess AGNs, respectively, in Smolčić et al. (2017b). By combining these classifications from the literature, An et al. (2021) excluded a total of 1916 AGNs from the 4509 VLA 3 GHz radio sources.

---

[2] www.eso.org/sci/observing/phase3/data_releases/uvista_dr2.pdf

[3] We refer the reader to An et al. (2021) for a detailed information on the selection criteria of the SFGs sample.

To ensure a clean sample of star-forming galaxies (SFGs), An et al. (2021) also excluded red galaxies with both ($M_{NUV} - M_r > 3.5$) and no detection in the Herschel bands (100, 160, 250, 350, and 500 μm) from their sample. A total of 761 galaxies meet these criteria, with 573 overlapping with the AGN sample. Consequently, after removing 2104 AGNs and/or red galaxies, a sample of 2405 SFGs was obtained with an SNR > 5 at VLA 3 GHz. From this sample, we selected those at redshifts $z > 1.5$ in order to construct mid-radio SEDs of galaxies at their rest frames. This reduces the sample size to 647 SFGs. Moreover, a robust SED analysis requires that all data points have at least an SNR > 1 leading to a sample of 189 galaxies at $1.5 < z < 3.5$. Lack of sensitive measurements at 0.3 and 0.6 GHz has led to this sample size reduction (An et al. 2021; Tisanić et al. 2019). As explained in Sect. 3.1, a few more galaxies have to be omitted from this analysis, resulting in a final sample size of 160 SFGs.

With an SFR > 100 $M_\odot$ yr$^{-1}$, the sample shows an increasing trend of SFR with stellar mass ($M_\star$) but with a large scatter (Fig. 1) that can refer to a mix of different SFG populations. According to Schreiber et al. (2015), the population of galaxies known as the star-forming main sequence (SFMS) obeys a universal log(SFR)–log($M_\star$) scaling relation. Figure 1 shows the expected SFMS relation at z= 1.5, 2.5, and 3.5. This relation indicates a roughly steady state star formation history. Galaxies with SFRs significantly above the SFMS (> 3× SFMS) are conventionally defined as starburst with intense star formation triggered by secular processes and/or mergers (see e.g., Speagle et al. 2014; Cooke et al. 2023, and references therein). Based on these definitions, the sample selected includes 85 starbursts and 75 SFMS SFGs.

## 3. REST-FRAME MID-RC SEDS

The selected sample at $1.5 < z < 3.5$ observed at 0.3 to 3 GHz allows a robust SED analysis at the rest-frame frequency range of $1 < \nu < 13.5$ GHz. This redshift (or rest-frame frequency) range is selected to avoid a probable curvature or flattening of the SEDs which is caused due to optically thick conditions or absorption effects (Condon 1992; Adebahr et al. 2013; Mulcahy et al. 2014; Marvil et al. 2015) as much as possible. At its rest frame, a galaxy emits the RC radiation at frequency $\nu_e$ with a flux density $S_{\nu_e}$ which is the sum of the free-free radiation of thermal electrons with flux density $S_{\nu_e}^{th}$ and the synchrotron emission of the nonthermal relativistic electrons with flux density $S_{\nu_e}^{nt}$:

$$S_{\nu_e} = S_{\nu_e}^{th} + S_{\nu_e}^{nt} = A_{th}\, \nu_e^{-\alpha_{th}} + A_{nt}\, \nu_e^{-\alpha_{nt}}, \quad (1)$$

with $\alpha_{th}$ and $\alpha_{nt}$ the thermal and nonthermal spectral indices, and $A_{th}$ and $A_{nt}$ constant scaling factors. The rest-frame frequency $\nu_e$ is connected to the observed frequency $\nu_o$ by $\nu_e = \nu_o(1+z)$. The SED model presented in Eq. 1 should be fitted by flux densities measured by an observer at frequencies $\nu_o$, i.e., $S_{\nu_o}$, hence, we construct a more practical expression. Taking into account the k-correction of $(1+z)^{-(1-\alpha)}$ for a power-law radio spectrum with index $\alpha$ ($S_\nu \propto \nu^{-\alpha}$), the luminosity is given by

$$L_{\nu_e} = \frac{4\pi D_L^2}{(1+z)^{(1-\alpha)}}\, S_{\nu_e}, \quad (2)$$

with $D_L$ the luminosity distance. Similar to flux density, the luminosity at the rest-frame frequency $\nu_e$ of a galaxy can be written as the sum of its thermal and non-thermal components:

$$L_{\nu_e} = L_{\nu_e}^{th} + L_{\nu_e}^{nt} \quad (3)$$

Applying Eq(2) for each thermal and nonthermal term in Eq(3), the luminosity at the rest-frame frequency $\nu_e$ of a galaxy is

$$L_{\nu_e} = \frac{4\pi D_L^2}{(1+z)^{(1-\alpha_{th})}}\, S_{\nu_e}^{th} + \frac{4\pi D_L^2}{(1+z)^{(1-\alpha_{nt})}}\, S_{\nu_e}^{nt}, \quad (4)$$

or using Eq(1),

$$L_{\nu_e} = \frac{4\pi D_L^2}{1+z} \left\{ \frac{A_{th}\, \nu_e^{-\alpha_{th}}}{(1+z)^{-\alpha_{th}}} + \frac{A_{nt}\, \nu_e^{-\alpha_{nt}}}{(1+z)^{-\alpha_{nt}}} \right\} \quad (5)$$

On the other hand, considering that $S_{\nu_e} = (\nu_e/\nu_o)^{-\alpha}\, S_{\nu_o}$, Eq(2) can be written in terms of the measured flux density $S_{\nu_0}$ as

$$L_{\nu_e} = \frac{4\pi D_L^2}{(1+z)}\, S_{\nu_o} \quad (6)$$

Eq(5) must be equal to and Eq(6), therefore,

$$S_{\nu_o} = \frac{A_{th}\, \nu_e^{-0.1}}{(1+z)^{-0.1}} + \frac{A_{nt}\, \nu_e^{-\alpha_{nt}}}{(1+z)^{-\alpha_{nt}}}, \quad (7)$$

in which the spectral index of the thermal emission $\alpha_{th} = 0.1$ for an optically thin free-free emission which is often the case in the mid-radio frequencies. To prevent dependencies on the units of the frequency space, Eq. (7) can be written as

$$S_{\nu_o} = \frac{A'_{th}}{(1+z)^{-0.1}} \left(\frac{\nu_e}{\nu_r}\right)^{-0.1} + \frac{A_{nt}\, \nu_r^{-\alpha_{nt}}}{(1+z)^{-\alpha_{nt}}} \left(\frac{\nu_e}{\nu_r}\right)^{-\alpha_{nt}}, \quad (8)$$

in which $\nu_r$ is the reference frequency and $A'_{th} = \nu_r^{-0.1}\, A_{th}$.

Following Tabatabaei et al. (2017), we use Bayesian Markov Chain Monte Carlo (MCMC) inference to fit the model given in Eq. (8) to the flux densities and derive the model parameters for the sample. This fitting method is favored over the chi-squared approach particularly when the noise has a non-Gaussian distribution which is often the case in astronomical observations. It is based on a wide library of models encompassing all plausible parameter combinations as priors and can provide robust statistical constraints on the fit parameters. To include all possible mechanisms of generation, acceleration, and cooling of cosmic-ray electrons, we take $\alpha_{\rm nt}$ to be uniformly distributed over the interval from 0 to -2.2 (see Tabatabaei et al. 2017; Longair 1994, and references therein). The scaling factors are sampled uniformly in the wide ranges $-1 < A'_{\rm th} < 1$, $1 < A_{\rm nt} < 30$ for flux densities in mJy. The negative values are included to just assess the robustness of the final results and the necessity for the thermal term. In case of a negative thermal term, the radio SED is fitted again using only a single nonthermal term with the thermal term set to 0.

We used the emcee code (Foreman-Mackey et al. 2013) to derive the posterior probability distribution function (PDF) for each of the free parameters $\alpha_{\rm nt}$, $A_{\rm th}$, and $A_{\rm nt}$. The medians of the posterior PDFs are reported as the final results with the 16% and 84% percentiles as uncertainties (Table A2). As examples, Fig. 2 shows the corner plots and posterior PDFs, and Fig. 3 the fitted SEDs of a few galaxies.

### 3.1. *MIGHTEE-COSMOS Radio SED Parameters*

As shown in Figs. 2 and 3, the MRC emission in the MIGHTEE-COSMOS SFGs can be described through two types of SEDs: One with both the thermal and nonthermal terms (top rows) and other with only a single nonthermal component (bottom rows). Only 11 galaxies show both the thermal and nonthermal components. In the rest of the sample, the thermal emission is below the uncertainty levels and a single nonthermal SED fits their observed fluxes well. The residuals are less than 25% at all frequencies for 75% of the sample. For the rest, the residuals exceed 50% at 0.3 and 0.6 GHz. It is likely that a curved SED fits better these sources but the sensitivity of their GMRT observations (SNR< 3) and the limited number of data points do not allow a precise treatment of a curved model. These sources, which will be investigated by adding sensitive LOFAR observations, are excluded in the current study reducing the sample size to a final sample of 160 galaxies (Fig. 1). While our analysis focuses on galaxies with robust SED fits, i.e., those having 5 data points with SNR> 1, possible effects of this selection are discussed in Appendix C.1.

The resulting $\alpha_{\rm nt}$ and thermal fraction at 1.3 GHz, $f^{\rm th}_{1.3} = S^{\rm th}_{1.3}/(S^{\rm th}_{1.3} + S^{\rm nt}_{1.3})$, together with their uncertainties are listed in Table A2. A histogram distribution of $\alpha_{\rm nt}$ is shown in Fig. 4-left for the entire sample. The nonthermal spectral index varies in the range $0.2 < \alpha_{\rm nt} < 1.3$ with a mean of $\alpha_{\rm nt} = 0.75$ and a standard error of $SE = \sigma/\sqrt{n} = \pm 0.01$ (with $\sigma$ the standard deviation and $n$ the sample size) over the sample. A slightly steeper spectrum is obtained for the SFMS galaxies (mean of $\alpha_{\rm nt} \simeq 0.8$). These are flatter than the average synchrotron spectral index of $\simeq 1$ obtained using similar SED analysis in samples of nearby starbursts (Galvin et al. 2018) and normal star-forming galaxies (Tabatabaei et al. 2017).

The thermal fraction of galaxies with both thermal and nonthermal components ranges from 10% to 35% with a mean of $17 \pm 8\%$ at rest frame 1.3 GHz which agrees with that of nearby galaxies at the same wavelength (Condon 1992; Tabatabaei et al. 2017). It also agrees with those predicted for analogs of nearby normal SFGs like M51 at high redshifts (Ghasemi-Nodehi et al. 2022).

The SFGs showing only a nonthermal MRC SED can refer to a population of galaxies different from the progenitors of nearby normal SFGs. The lack of thermal emission was also reported by Thomson et al. (2019) for a sample of submillimeter galaxies, though with a steep $1.4 < \nu < 6$ GHz spectrum ($\alpha^{\rm 6GHz}_{\rm 1.4GHz} = 1.06 \pm 0.04$). For a sub-sample of the MIGHTEE-COSMOS SFGs at $z > 1.5$, we used a de-reddened H$\alpha$ emission as a tracer of the thermal free-free emission to separate the thermal and nonthermal radio components independently from the radio SED modeling (see Appendix B). In most cases, the resulting thermal fractions are negligible and agree with our SED results within the uncertainties. In general, an excess of the nonthermal emission (or weakness of the thermal emission) is expected in highly star-forming galaxies if the nonthermal emission increases super-linearly, while the thermal emission increases linearly with the SFR as predicted by Tabatabaei et al. (2017) based on studies in nearby galaxies. This is further investigated for the MIGHTEE-COSMOS high-z SFGs in Sect. 6.

## 4. MRC LUMINOSITY

To obtain the total energy output of the high-z SFGs emitting in mid-radio frequencies, we integrate their SEDs over $1 - 10$ GHz. This is useful to compare the energy budget of galaxies with redshift. Moreover, it also allows us to study the energy balance between the radio and infrared domains. Following (Tabatabaei et al. 2017), the total energy output of galaxies emitting over 1-10 GHz frequency range is given by the integrated MRC luminosity as

$$\mathrm{MRC} = \int_{1\,\mathrm{GHz}}^{10\,\mathrm{GHz}} L_{\nu_e}\, d\nu_e, \quad (9)$$

with $L_{\nu_e}$ given by Eq.(6) and $S_{\nu_o}$ by Eq.(8). The integrated MRC luminosities obtained as well as the monochromatic flux densities of the sample are listed in Table A2. Figure 5 shows a histogram of the MRC in a logarithmic scale. It varies between $10^{6.4}\,L_\odot$ and $10^8\,L_\odot$ with a mean of $10^7\,L_\odot$ in the sample which is 3 orders of magnitude larger than that found in the KINGFISH sample of nearby galaxies (Tabatabaei et al. 2017).

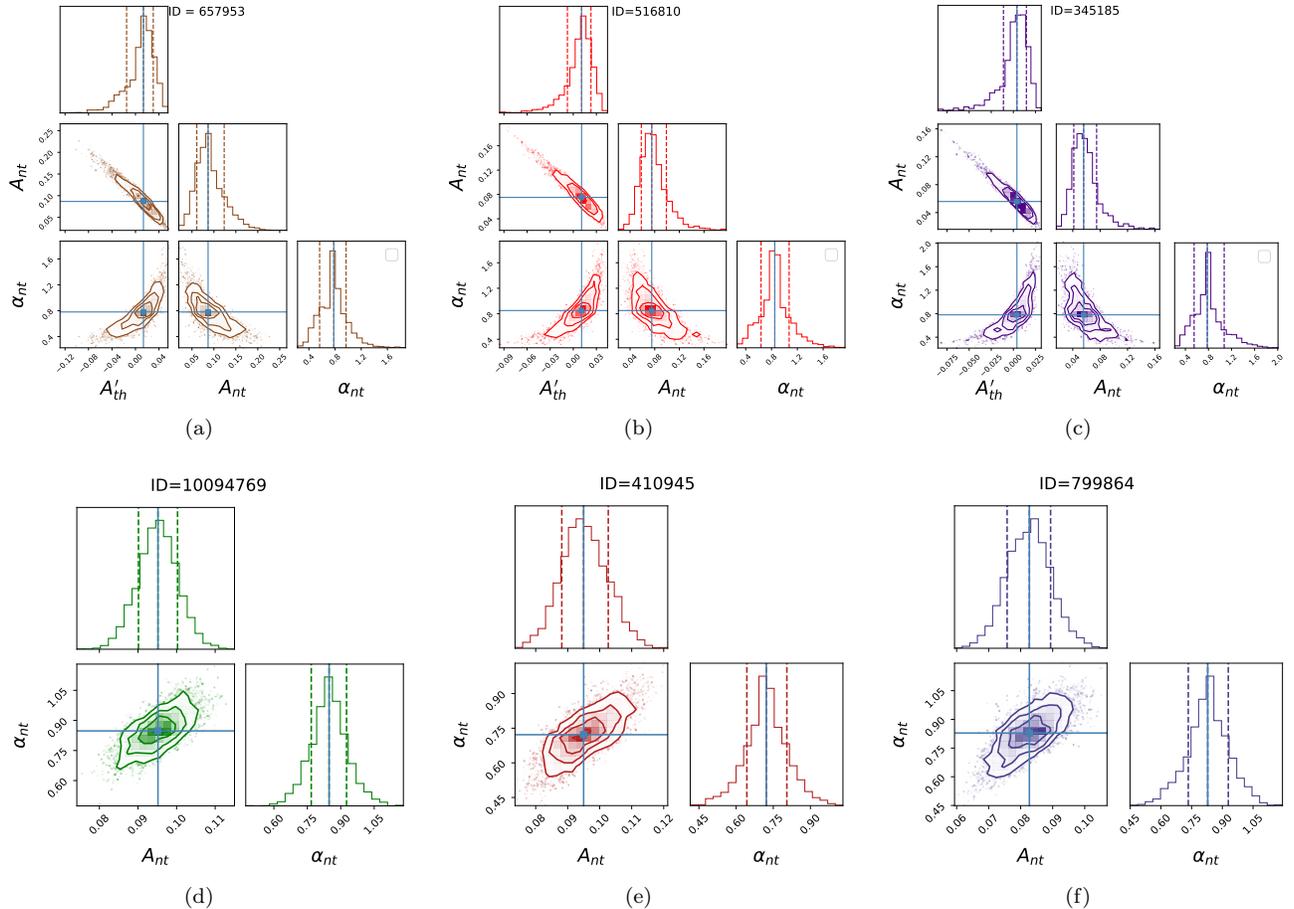

Figure 2: Corner plots showing the two-dimensional posterior probability distribution function (PDF) of $A_{th}$, $A_{nt}$ and $\alpha_{nt}$ fitted to the rest-frame SEDs and their 0.16, 0.5, and 0.84 percentiles. The best-fitting median value $\mu$ is at the $50^{th}$ indicated by the blue squares and solid lines. Within one standard deviation $\sigma$ of the median value $\mu$, 68% of the density distribution is contained; 34% of the density is above $\mu$ but below $\mu + \sigma$ corresponds to the $84^{th}$ percentile, and 34% of the density is below $\mu$ but above $\mu - \sigma$ corresponds to the $16^{th}$ percentile.

Multi-band cosmological RC surveys are required to obtain the MRC luminosity by integrating the radio SEDs. If such multi-band surveys are unavailable, using a calibration relation can be a suitable alternative. By using the 1.3 GHz and 3 GHz rest frame luminosities, the following calibration relation is found to derive the integrated MRC luminosity over the redshift range of $1.5 < z < 3.5$.

$$\frac{\mathrm{MRC}}{\mathrm{erg\,s^{-1}}} = a\left(\frac{\nu_{1.3}\,L_{1.3}}{\mathrm{erg\,s^{-1}}}\right) + b\left(\frac{\nu_3\,L_3}{\mathrm{erg\,s^{-1}}}\right), \quad (10)$$

with $a = -0.62$ and $b = 2.89$. This relation connects the observed fluxes at $1.3\,\mathrm{GHz}/(1+z)$ and $3\,\mathrm{GHz}/(1+z)$ to the MRC luminosity of SFGs at redshift $z$ using Eq.(6).

Surveys such as the on-going MIGHTEE S-band and L-band observations can be combined using the above expression to obtain the MRC luminosity at $z = 0$. At higher redshifts low-frequency observations with e.g., LOFAR or SKA-LOW can be used to study the energy budget of galaxies emitting in mid-radio.

5. EQUIPARTITION MAGNETIC FIELD

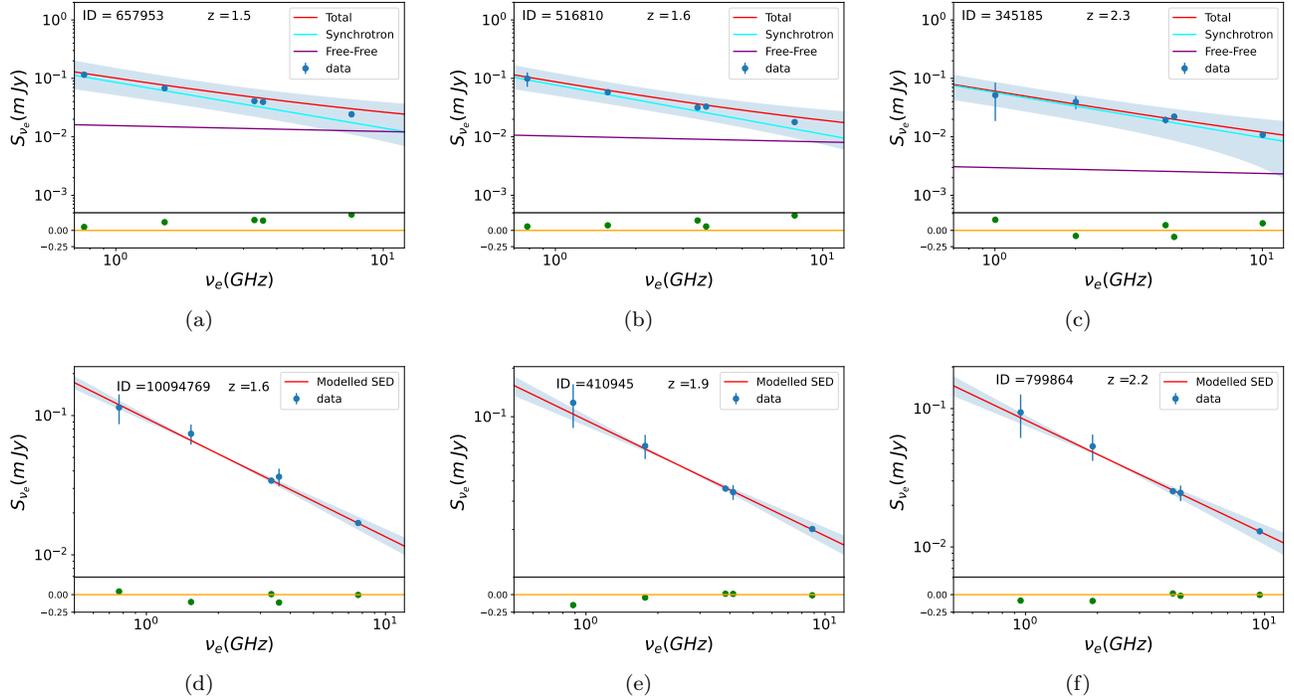

Figure 3: Radio continuum SEDs at rest-frame (rest-frame flux vs. rest-frame frequency) and their 84% confidence bounds (shaded area) for 6 selected galaxies in the MIGHTEE-COSMOS field. Upper panels show examples with both thermal and nonthermal components while lower panels have only a single nonthermal component. The relative residuals (modeled-observed/observed ratio) are shown at the bottom of each plot.

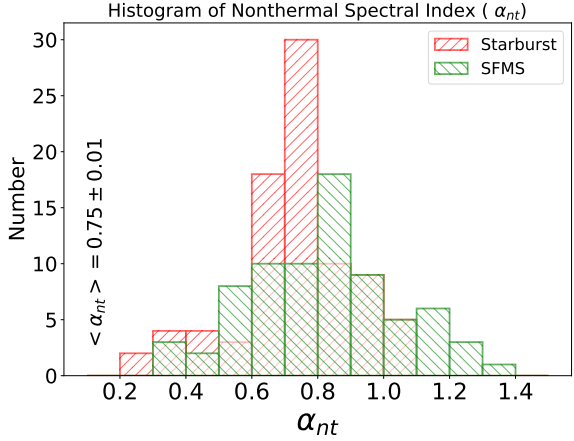

Figure 4: Histogram distribution of the non-thermal spectral index $\alpha_{nt}$ of the SFGs (starburst and SFMS galaxies) in the redshift range $1.5 < z < 3.5$.

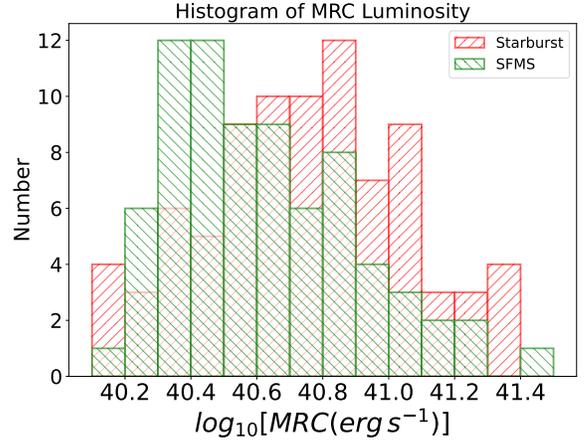

Figure 5: Histogram distribution of the rest-frame MRC luminosity of the SFGs (starbursts and SFMS galaxies) in the redshift range $1.5 < z < 3.5$.

The strength of the magnetic field in high-z galaxies and its evolution over cosmic time are pressing questions. As explained in Appendix D the synchrotron luminosity $L_{nt}$ of a galaxy can be used to estimate its magnetic field strength B using the following expression,

$$B = B_0 \left(\frac{\cos(i)}{\cos(i_0)}\right)^{-1/4} \left(\frac{L_{nt}}{L_{nt0}}\right)^{1/4} \left(\frac{M_\star}{M_{\star 0}}\right)^{-0.1}, \quad (11)$$

with $M_\star$ the stellar mass and $i$ the inclination angle. Index '0' in the above expression refers to a reference galaxy. We opt to use a highly star forming galaxy with known magnetic field strength as the reference galaxy. NGC 253 is an ideal reference: Its core resembles starburst galaxies while its disk has properties of SFMS galaxies. At a distance of 3.5 Mpc, NGC 253 has a stellar mass of $M_{\star 0} \simeq 10^{10.98}$ (Karachentsev et al. 2021) an inclination of $i_0 = 78°$ and a flux density of $S_0 = 6.22 \pm 0.07$ Jy at 1.3 GHz (Williams & Bower 2010). The strength of its total equipartition magnetic field estimated by Heesen et al. (2014) is $B_0 = 46 \pm 10\,\mu G$. As the inclinations of high-z galaxies are unknown, we assume that the inclination angles are randomly distributed over $i = 0$ to $i = \pi/2$. This leads to a mean value of $\left\langle \left( \frac{\cos(i)}{\cos(i_0)} \right)^{-1/4} \right\rangle \simeq 1$ over the sample. Using this in Eq.(11), results in a magnetic field strength changing from 91 to 192 $\mu G$ (Fig. 6) with a mean of 135 $\mu G$ (155 $\mu G$ for starbursts) and statistical error of 1 $\mu G$. We however, note that the change in $i$ results in $\leq 0.4$ dex uncertainty in the magnetic field estimation.

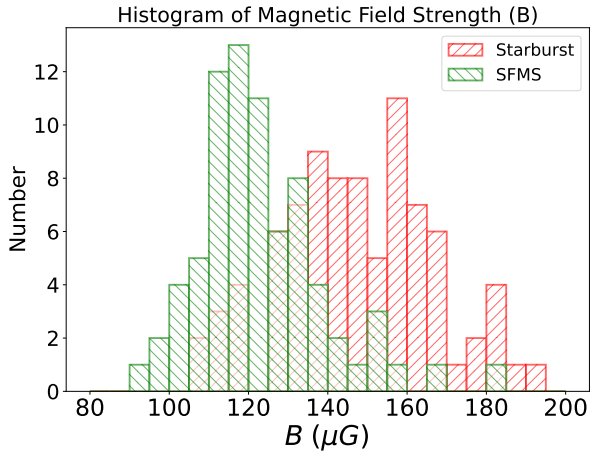

Figure 6: A histogram distribution of the equipartition magnetic field strength in the MIGHTEE-COSMOS SFGs (starburst and SFMS galaxies) in the redshift range of $1.5 < z < 3.5$.

The resulting mean B strength in the MIGHTEE-COSMOS SFGs at $1.5 < z < 3.5$ is larger than that found for the KINGFISH sample of nearby galaxies ($13.5 \pm 5.5\mu G$, Tabatabaei et al. 2017) by about one order of magnitude taking into account the uncertainties. Possible reasons such as an excess amplification due to star formation activities are discussed in Sect. 8.

## 6. DISCUSSION

In previous sections, we presented the radio SEDs of the SFGs at $1.5 < z < 3.5$, obtained their MRC luminosity, and estimated their magnetic field strength. In this section, we investigate any variations with redshift and possible connections with the SFR traced with the total IR luminosity (TIR). Then, taking into account the variation in the radio spectral index (and k-correction), the IRRC is revisited in the COSMOS-MIGHTEE SFGs. Finally, the SFR is calibrated directly in radio at high-z and the corresponding relations are presented using both the monochromatic RC and integrated MRC luminosities.

### 6.1. Evolution of Radio SEDs and Effect of Star Formation Activity

As shown in Sect. 3.1, in the redshift range of $1.5 < z < 3.5$, the nonthermal synchrotron spectrum is on average flatter ($\alpha_{\rm nt} = 0.75 \pm 0.01$) than that of both starbursts ($\alpha_{\rm nt} \simeq 1$, Galvin et al. 2018) and normal star forming galaxies ($\alpha_{\rm nt} = 0.97 \pm 0.03$, with error the standard error, Tabatabaei et al. 2017) in local universe measured using the same technique. Does this difference imply a cosmic evolution of the SED shape?

In Fig. 7, we plot $\alpha_{\rm nt}$ vs redshift for the MIGHTEE-COSMOS SFGs. Although the scatter is high, the mean $\alpha_{nt}$ values in redshift bins of 0.2 (red points in Fig. 7) show a trend of flattening with redshift.

With a Pearson correlation coefficient of $r_p = -0.9$, the mean $\alpha_{nt}$ values are best described by the following power-law relation in terms of $(1 + z)$:

$$\alpha_{nt} = (1.8 \pm 0.1) \times (1 + z)^{(-0.8 \pm 0.1)}. \quad (12)$$

Accordingly, the mean nonthermal spectral index flattens by about 37% from z=1.5 to z=3.5. This implies that the shape of the radio SEDs of the SFGs varies with $z$ being flatter at earlier cosmic times. This is in contrast with the assumption of an invariant SED shape which is often used as the basis for considering a standard k-correction with a fixed spectral index (e.g., Delhaize et al. 2017; Tisanić et al. 2019). Using the VLA observations of the $\mu$Jy star-forming radio sources in the GOODS-N field, Murphy et al. (2017) also found a flatter spectrum at high redshifts ($\alpha_{nt} \simeq 0.61$ at $<z> = 1.24 \pm 0.15$).

Resolved studies in normal star-forming galaxies show that the synchrotron spectrum is flatter and closer to the injection spectrum of cosmic ray electrons in star-forming regions (Nasirzadeh et al. 2024; Tabatabaei et al. 2022; Hassani et al. 2022; Tabatabaei et al. 2013b). Similarly, it has been shown that $\alpha_{\rm nt}$ flattens with increasing star formation rate surface density, $\Sigma_{\rm SFR}$, in the KINGFISH sample globally (Tabatabaei et al. 2007). Simulating the RC emission from analogs of nearby main sequence galaxies back to $z = 3$, Ghasemi-Nodehi et al. (2022) predicted a flattening of $\alpha_{\rm nt}$ with redshift. They linked this to the cosmic evolution of the SFR (Schreiber et al. 2015) and the possibility of a flattening of $\alpha_{\rm nt}$ with $\Sigma_{\rm SFR}$. The cosmic ray electron population can be younger and more energetic in highly star-forming galaxies. Does the evolution of massive star formation play a role in the evolution of $\alpha_{\rm nt}$ observed in the MIGHTEE-COSMOS SFGs?

It is worth mentioning that, in starburst galaxies, in addition to the star formation activity, other factors might also lead to an observed flattening of the nonthermal spectrum such as absorption due to an optically thick ISM (Murphy et al. 2013; Barcos-Muñoz et al. 2017) and other cooling mechanisms of the cosmic ray electrons such as bremsstrahlung and ionization losses (Lacki et al. 2010). In the case of the RC absorption in an optically thick region, a lack of the RC emission is expected compared to the FIR emission. The ionization loss is more effective for low-energy cosmic rays ($\leq 1.3\,{\rm GeV}$, e.g., Murphy 2009) and at low frequencies as its time scale is $\tau_{\rm ion} \simeq \nu^{0.5}$. Similar to the RC absorption in an optically thick region, the loss mechanisms other than the synchrotron loss lead to a suppression of the nonthermal radio emission compared to the IR emission. We will study the IRRC in our sample of SFGs in Sect. 6.3.

The impact of star formation on the rest-frame mid-RC SEDs can be investigated by cross-correlating $\alpha_{\rm nt}$ with the SFR per unit area. As information on the size of the sources is unavailable, the SFR per unit mass (sSFR = SFR/$M_\star$) is used instead. As shown in Fig. 7-(b), we find that the nonthermal spectrum becomes flatter with sSFR following

$$\alpha_{\rm nt} = (-0.25 \pm 0.02) \times \log_{10}\left(\frac{\rm sSFR}{\rm yr^{-1}}\right) - (1.37 \pm 0.18) \quad (13)$$

This relation is very similar to that found between $\alpha_{\rm nt}$ and $\Sigma_{\rm SFR}$ of the normal star forming galaxies (Tabatabaei et al. 2017). This agreement between the MIGHTEE-COSMOS high-z/starbursts and the KINGFISH nearby/normal SFGs indicates that the same astrophysical phenomena must be at work irrespective of their possibly different ISM conditions: A more energetic population of cosmic ray electrons are produced due to the SFR activity per unit area or per unit galaxy mass. Supernova explosions can accelerate particles to very high energies. This can lead to synchrotron radiation with a flat nonthermal spectral index. In starburst galaxies, there is a high rate of star formation, which can lead to the production of supernovae and the acceleration of high-energy particles. This can also explain the flatter synchrotron spectral index observed in these starburst galaxies at $1.5 < z < 3.5$ compared to the nearby less star-forming galaxies.

Including the RC data at frequencies lower than 1 GHz, (Niklas et al. 1997) modeled the radio SEDs of 74 nearby galaxies. They obtained a somewhat flatter spectral index $\alpha_{\rm nt} = 0.83 \pm 0.02$ than that found in the 1-10 GHz range by Tabatabaei et al. (2017) which is expected because of possible low-frequency flattening. On the other hand, Niklas et al. (1997) also found indications for flattening of the nonthermal spectrum in galaxies with strong star formation activity. For a sample of local luminous and ultra-luminous IR galaxies, Murphy et al. (2013) found that the total observed spectral index (measured using the observed fluxes at 1.4 and 8.4 GHz) flattens with increasing sSFR linking it to absorption due to an optically thick condition of the ISM of starburst galaxies. On the other hand, accounting for the absorption effects, Galvin et al. (2018) showed that although the curvatures can occur in the SEDs of the starbursts, their pure synchrotron spectra are not flatter than those of normal star-forming galaxies (Tabatabaei et al. 2017).

To determine the extent to which the cosmic evolution of the nonthermal spectral index occurs independently of the specific star formation rate, we calculate the partial rank correlation $\rho_{23,1}$,

$$\rho_{23,1} = \frac{\rho_{23} - \rho_{12}\rho_{13}}{(1-\rho_{12}^2)^{1/2}(1-\rho_{13}^2)^{1/2}}, \quad (14)$$

which evaluates the correlation between parameters 2 and 3 when parameter 1 is kept as constant. In Eq.(14), $\rho$ is the Spearman rank correlation coefficient evaluating the strength and direction of the correlations. It takes a value in the range of -1 (for a perfect negative or inverse relationship) to 1 (for a perfect positive or direct relationship), and the value 0 means that there is no linear relationship. Taking the sSFR as the control parameter 1, $\alpha_{nt}$ as parameter 2, and $z$ as parameter 3, we obtain a weak partial rank of $\rho_{\alpha_{\rm nt} z,{\rm sSFR}} = -0.18$ for the MIGHTEE-COSMOS SFGs. This indicates that once the influence of the control variable sSFR is removed, the two variables $\alpha_{nt}$ and $z$ do not show a significant association. Hence, it is deduced that the cosmic evolution of $\alpha_{\rm nt}$ is mainly due to the evolution in massive star formation activities.

### 6.2. Cosmic Evolution of Magnetic Field and Role of Star Formation

As shown in Sect 5, the magnetic field of the MIGHTEE-COSMOS SFGs is, on average, stronger than that of the nearby galaxies. We further discuss any systematic variation of B with redshift. Figure 6-right, shows a monotonic increase of B over the redshift interval $1.5 < z < 3.5$. The best curve fit results in the following relation with a Pearson correlation coefficient $r_p = 0.5$,

$$\left(\frac{\rm B}{\mu \rm G}\right) = (55 \pm 7) \times (1+z)^{(0.7 \pm 0.1)}. \quad (15)$$

This evolution can be due to a systematic change in the origin and amplification mechanism of the magnetic field over cosmic time. Generally, the origin of the magnetic field in galaxies is still a matter of debate (see e.g., Widrow 2002; Subramanian 2016, for a review). The mean field dynamo mechanism proposed to explain the ordered coherent field in galaxies (e.g., Krause & Raedler 1980) is challenged by the fact that the magnetic field must have been already saturated at an early cosmic time due to quenching of the large-scale dynamo action (e.g., Subramanian & Brandenburg 2004; Vishniac & Cho 2001). On the other hand, observations of the nearby galaxies show that the large-scale ordered magnetic field traced with the synchrotron polarization is scaled with rotation speed and dynamical mass of galaxies (Tabatabaei et al. 2016). Similar correlations are found by the TNG50 cosmological simulations (Hosseinirad et al. 2023) taking into account the mean magnetic field strength. This can indicate that a major part of the ordered field in galaxies is generated by the B-gas coupling and magnetic flux freezing (Tabatabaei et al. 2016). Observations also show that the total field strength, traced by the synchrotron total power, is scaled with the SFR in nearby galaxies (Tabatabaei et al. 2017; Heesen et al. 2014; Chyży et al. 2011). This has been linked to a small-scale dynamo mechanism that occurs in star-forming regions (Tabatabaei et al. 2022). Star formation activities and supernova shocks amplify the magnetic field through a small-scale dynamo effect. Theoretically, a power-law relation between B and SFR with an index of 0.3 (B $\propto$ SFR$^{0.3}$) is expected if the magnetic field is amplified due to supernova-injected turbulence (Schleicher & Beck 2013). Investigating the total magnetic field B vs the SFR for the MIGHTEE-COSMOS SFGs, we find a tight correlation between these two quantities (Fig. 8). A linear regression fit leads to the following relation,

$$\left(\frac{B}{\mu G}\right) = 10^{(1.5 \pm 0.1)} \times \left(\frac{SFR}{M_\odot\, yr^{-1}}\right)^{(0.25\,\pm\,0.05)} \quad (16)$$

With a power-law index of 0.25±0.05, the B–SFR correlation in the MIGHTEE-COSMOS SFG sample at $1.5 < z < 3.5$ agrees with the theory of the magnetic field amplification due to a turbulent dynamo in star-forming regions. As the same relation holds in the nearby galaxies (e.g., Tabatabaei et al. 2017), the small-scale dynamo must have always been functioning in galaxies, however, its intensity has dropped over cosmic time due to the cosmic evolution of the SFR (e.g., Schreiber et al. 2015) resulting in the weakening of B with time (or strengthening of B with redshift).

We assess the influence of the SFR in the cosmic evolution of the magnetic field using the partial rank correlation defined in Eq.(14). The correlation $\rho_{23,1}$ is calculated with "1" representing the SFR as the control parameter, "2" the magnetic field strength B, and "3" the redshift $z$. For MIGHTEE-COSMOS SFGs, this correlation is weak, $\rho_{Bz,SFR} \simeq 0.2$ and hence, the SFR is the main cause of the increase in the magnetic field strength with redshift.

We note that the mean magnetic field strength of $\simeq 135\,\mu G$ agrees with the local starburst estimates of Lacki & Beck (2013). However, this estimate is likely a lower limit to the magnetic field strength at high-z considering their much higher SFR. A more realistic calibration should use a reference galaxy at similar redshifts ($1.5 < z < 3.5$) that has been unavailable for this study.

### 6.3. Revisiting The IR-Radio Correlation

The tight correlation between the IR and RC emission (IRRC) of star-forming galaxies (e.g., Helou et al. 1985; de Jong et al. 1985; Yun et al. 2001) has been used to trace the SFR in galaxies (e.g., Bell 2003b; Murphy et al. 2011) and to distinguish between AGNs and SFGs at high-z (Ibar et al. 2008), although with a large scatter (Jarvis et al. 2010; Morić et al. 2010; Sargent et al. 2010; Sargent et al. 2010; Appleton et al. 2004; Bourne et al. 2010). However, these are complicated by the observed change in correlation with redshift (e.g., Seymour et al. 2009; Ivison et al. 2010; Ivison et al. 2010; Basu et al. 2015; Magnelli et al. 2015; Delhaize et al. 2017). Recent work also suggests that the apparent redshift evolution can be due to a dependence on stellar mass (e.g., Delvecchio et al. 2021b). The possibility that this variation can be rooted in the assumed k-correction (Delhaize et al. 2017; An et al. 2021) motivates us to reinvestigate this correlation in the COSMOS field taking into account the variation of the SEDs obtained in this work. We characterize the IRRC by the $q$ parameter defined as the logarithmic ratio of the total infrared luminosity (8 − 1000 $\mu m$; $L_{TIR}$) to the 1.3 GHz RC luminosity ($L_{1.3}$):

$$q = \log_{10}\left[\frac{L_{TIR}}{3.75 \times 10^{12}\,W} \times \frac{W\,Hz^{-1}}{L_{1.3}}\right]. \quad (17)$$

The k-correction used to obtain $L_{1.3}$ is based on the variable spectral index obtained in Sect. 3. The resulting q changes between 1.6 and 2.8 (above the AGN dominated values, Algera et al. 2020; Klutse et al. 2024) with a mean value of $q = 2.2 \pm 0.01$ over $1.5 < z < 3.5$. This is lower than that reported for local galaxies including ultra-luminous infrared galaxies (ULIRGS, $q = 2.64 \pm 0.02$, Bell 2003b). Moreover, no systematic trend is found for $q$ against redshift. Fig. 9-a shows that the dispersion around the fitted $q = 2.2$ is about 0.2 dex. Thus, the correlation between the IR and the nonthermal RC emission remains invariant over the redshift range $1.5 < z < 3.5$. This agrees with Algera et al. (2020), who obtained the same $q$ with no $z$-dependency for their sub-millimeter selected sample of SMGs over the same redshift range. On the other hand, they concluded that the evolution of the IRRC reported by other authors can be due to sample selection in the radio regime. However, using a radio-based selected sample, we also obtained no IRRC evolution at $1.5 < z < 3.5$.

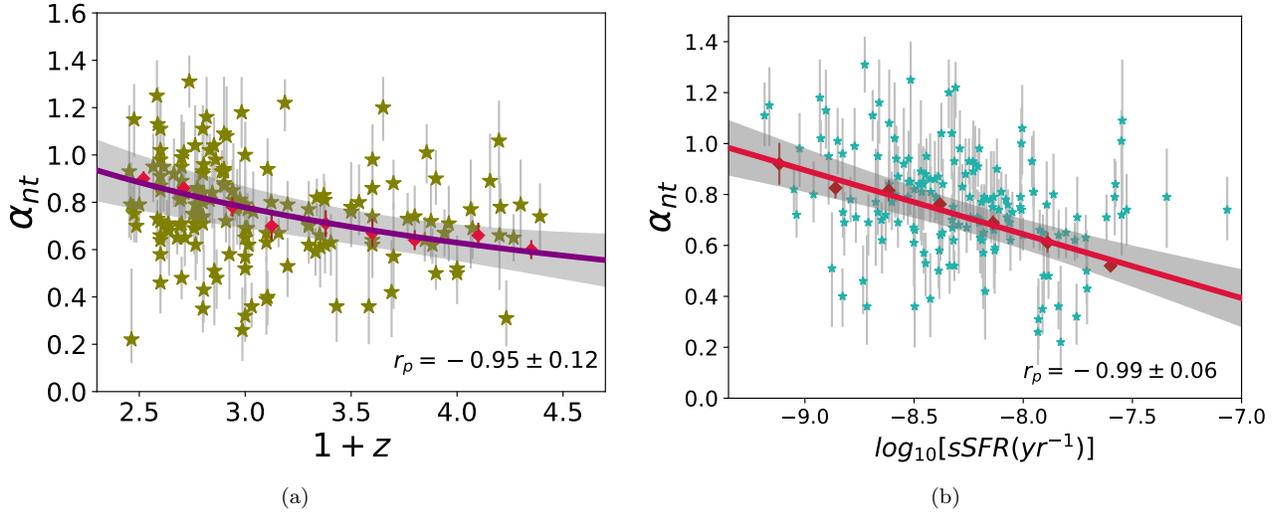

Figure 7: Nonthermal spectral index $\alpha_{\rm nt}$ against redshift $z$ ($a$) and specific star formation rate sSFR ($b$) for SFGs located at $1.5 < z < 3.5$. Curves show best fits to the mean values in redshift bins of 0.2 (red points) with their Pearson correlation coefficients $r_p$. Gray shaded areas show 99% confidence bounds in both panels.

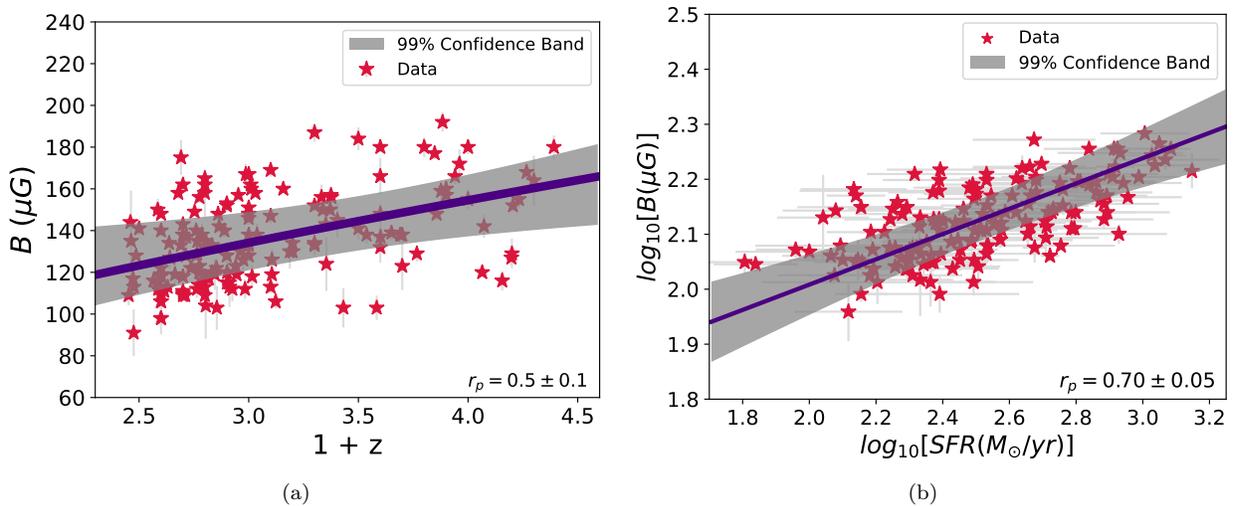

Figure 8: Cosmic evolution of the magnetic field strength B of the SFGs at $1.5 < z < 3.5$ ($a$). Also shown is B against the SFR in the logarithmic scale ($b$). Curves show the best fits to all data points with $r_p$ representing their corresponding Pearson correlation coefficient, and gray shaded areas the 99% confidence bounds.

We further investigate the effect of a fixed $k$-correction in the evolution of the IRRC by assuming that the radio spectral index is fixed and equal to the mean of its distribution ($\alpha_{\rm nt} = 0.75$). The resulting $q$ shows a shallow decrease with redshift (Fig. 9-b) $q = (2.7 \pm 0.1) \times (1+z)^{(-0.16 \pm 0.04)}$, which is in agreement with that reported by (Tisanić et al. 2019; Molnár et al. 2018; Delhaize et al. 2017; Magnelli et al. 2015). Hence, the reported evolution in the IRRC is caused by the variation in the rest-frame SEDs and the scatter in $\alpha_{\rm nt}$ if they are not taken into account.

It should be emphasized that the relatively low value of $q = 2.2$ at $1.5 < z < 3.5$ still agrees with the mean values found by Tisanić et al. (2019); Molnár et al. (2018); Delhaize et al. (2017); Magnelli et al. (2015) in the same redshift range. In a separate paper, we will address the IRRC at lower redshifts ($z < 1.5$) after determining $\alpha_{\rm nt}$ taking into account possible low-frequency ($\nu < 1$) curvatures in rest-frame SEDs of the MIGHTEE-COSMOS SFGs.

The IRRC correlation defined by Eq(17) uses the total IR (TIR) luminosity while referring to a single monochromatic radio luminosity. In other words, the IR SED but not the RC SED is involved studying the correlation between the IR and RC domains. Following Tabatabaei et al. (2017), evaluating the ratio of TIR to integrated MRC luminosities,

$$q_{\rm MRC} = \log_{10}\left(\frac{\rm TIR}{10^3 \, \rm MRC}\right), \quad (18)$$

is physically more justified and insightful as it can also provide a measure for the balance of the energy budget of galaxies emitting in the IR and Mid-RC regimes. Figure 9-c shows that $q_{\rm MRC}$ remains invariant with redshift taking into account the variation of $\alpha_{\rm nt}$. This is similar to what we found for the q parameter but with a slightly smaller scatter around the mean value ($q_{\rm MRC} = 2.2$). Hence, over the redshift interval of $1.5 < z < 3.5$, no systematic change in the total energy output of SFGs emitting in IR relative to that emitting in radio has occurred.

We also explored any correlation between q (and $q_{\rm MRC}$) and other parameters such as the SFR, magnetic field strength, and stellar mass ($M_\star$). Figure 10 shows that none of those parameters are correlated with q (or $q_{\rm MRC}$, not shown). The lack of correlation between q and SFR agrees with other global studies (Yun et al. 2001). On the other hand, Delvecchio et al. (2021c) found an indication for a change in the IRRC correlation with the stellar mass for $z < 3.5$ galaxies which disagrees with our result over $1.5 < z < 3.5$. If the correlation found by Delvecchio et al. (2021c) is mainly due to the lower redshifts ($z < 1.5$) will be addressed in a separate study.

As discussed in Sect. 6.1, an excess of the IR emission is expected with respect to the RC luminosity if the medium is optically thick to the RC emission or in case of a dominant loss process different from the synchrotron loss. However, this is not confirmed by the values obtained for q and $q_{\rm MRC}$. Hence, the flattening of $\alpha_{\rm nt}$ with sSFR observed in the COSMOS-MIGHTEE SFGs (Sect. 6.1) cannot be due to absorption in an optically thick ISM or the ionization loss of cosmic ray electrons. Lacki et al. (2010) proposed that the synchrotron of secondary cosmic ray electrons and positrons produced due to the ionization loss of primary cosmic rays is responsible for a linear IRRC in starbursts. However, the IRRC observed in these high-z SFGs deviates from linearity (see also Sect. 6.4).

The SFGs being more luminous in mid-radio than IR, compared to those in the local universe, must have the same root in their $\alpha_{\rm nt}$ being flatter (i.e., more energetic cosmic rays, Sect. 6.1) and the magnetic fields becoming stronger (Sect. 6.2) at high redshifts. These, both, should lead to an excess of the RC luminosity with redshift and a drop in q or $q_{\rm MRC}$ as obtained. As such, a sub-linear correlation for the TIR (as SFR tracer) against MRC (and monochromatic RC) luminosities is expected (see Sect. 6.4).

### 6.4. SFR Calibrations at High-z

Measuring the SFR is the first step in understanding the evolution and quenching of galaxies over cosmic time. In Sect. 6.3, we showed that the IRRC is invariant at $1.5 < z < 3.5$ if variations in the radio SED and $\alpha_{\rm nt}$ are taken into account, although q differs from that in the local universe. As follows, we explore the SFR calibration relations using two methods: 1) cross-correlating the SFR with the RC luminosities provided that the reference SFR is given by the TIR luminosity and 2) using physically motivated radio-based calibrations to obtain $\rm SFR_{RC}$ and then comparing it with $\rm SFR_{TIR}$. In both methods, we investigate the usage of monochromatic and integrated MRC luminosities and consider the radio SED variations.

As for the first method, the SFR traced using the TIR luminosity, $\rm SFR_{TIR}$, is plotted against the monochromatic RC luminosities at 1.3 and 3 GHz ($\nu L_\nu$) in Fig. 11. In order to investigate possible variation in the SFR calibration with redshift over $1.5 < z < 3.5$, the correlations are obtained separately over the two ranges of $1.5 < z < 2$ and $2 < z < 3.5$. The Pearson correlation coefficients and the fitted slopes agree within the errors in the two redshift ranges and the correlations are all sub-linear (see Table 1).

For the entire redshift range $1.5 < z < 3.5$, the following calibration relations are found,

$$\left(\frac{\rm SFR}{\rm M_\odot \, yr^{-1}}\right) = 10^{(-28.9 \pm 1.7)} \times \left(\frac{\nu_{1.3} \, L_{\nu_{1.3}}}{\rm erg \, s^{-1}}\right)^{(0.78 \pm 0.04)}, \quad (19)$$

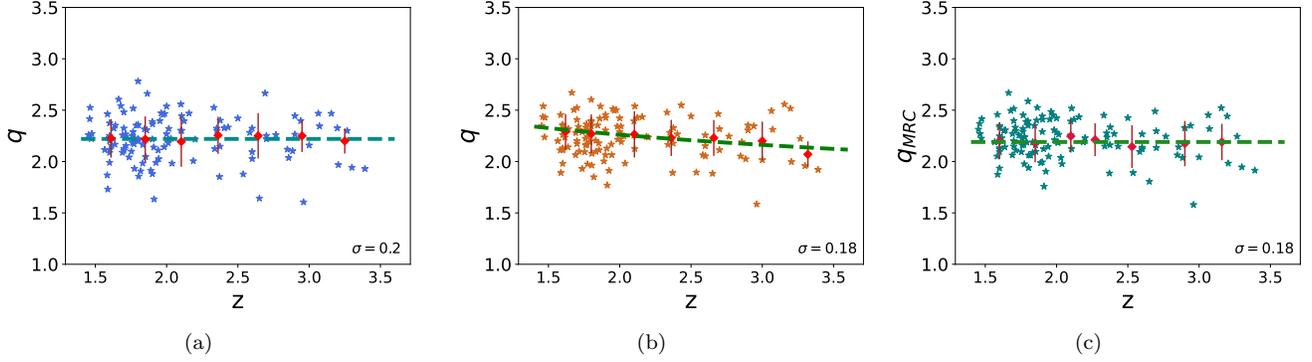

Figure 9: The IRRC correlation ($q$-parameter) obtained using the variable $\alpha_{\rm nt}$ (a) or assuming a fixed $\alpha_{nt} = 0.7$ (b) against redshift $z$. Panel (c) is the same as panel (a) but with the MRC used instead of the monochromatic radio luminosity at 1.3 GHz ($q_{\rm MRC}$). Dashed lines show the best fits to average values in redshift bins of 0.2 representing $q = 2.2 \pm 0.1$ (a), $q \simeq (1+z)^{(-0.16\pm 0.03)}$ (b), and $q_{\rm MRC} = 2.2 \pm 0.1$ (c). In all panels $\sigma$ shows the scatter from the fits.

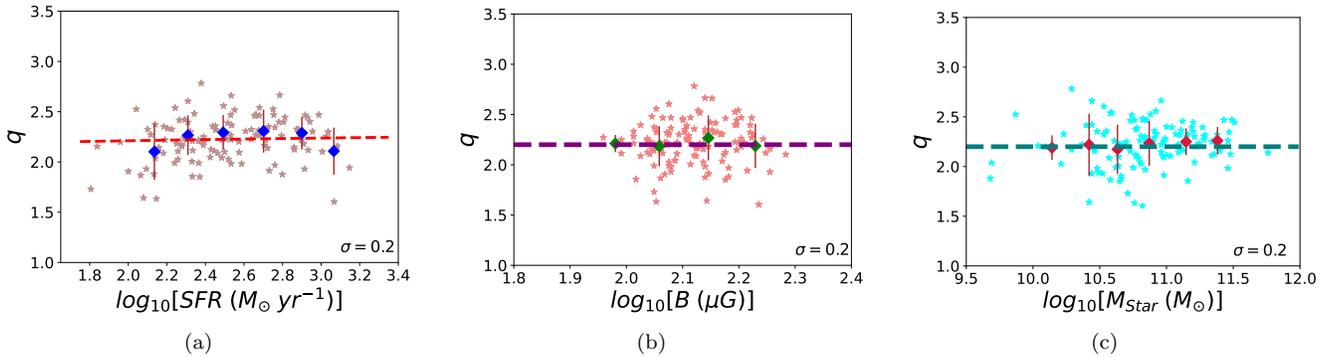

Figure 10: The IRRC correlation (q-parameter) against SFR (a), magnetic field (b), and stellar mass (c). The dashed line shows $q = 2.2$ in all panels.

$$\left(\frac{\rm SFR}{\rm M_\odot\,yr^{-1}}\right) = 10^{(-30.5\,\pm\,1.5)} \times \left(\frac{\nu_3\,L_{\nu_3}}{\rm erg\,s^{-1}}\right)^{(0.82\,\pm\,0.03)}, \quad (20)$$

in which SFR$_{\rm TIR}$ is replaced by SFR.

The correlation of the SFR (traced with TIR) is tighter with the MRC than the monochromatic luminosities (see Table 1). We find the following calibration relation using the MRC luminosity for the SFGs located at $1.5 < z < 3.5$:

$$\left(\frac{\rm SFR}{\rm M_\odot\,yr^{-1}}\right) = 10^{(-30.7\,\pm\,1.4)} \times \left(\frac{\rm MRC}{\rm erg\,s^{-1}}\right)^{(0.83\,\pm\,0.03)}, \quad (21)$$

with a dispersion of $\simeq 0.10$ dex which is smaller than the monochromatic calibrations (>0.16 dex, Fig. 11). We also note that a better agreement holds between the two redshift ranges using the MRC luminosity.

Table 1 shows more detailed information on the SFR calibrations given by $\log_{10}({\rm SFR}) = {\rm a} + {\rm b}\,\log_{10}({\rm X})$, with X the radio luminosity.

Table 1: Correlation relations between SFR and RC luminosities ($\log_{10}({\rm SFR}) = {\rm b}\,\log_{10}({\rm L}_{\rm RC}) + {\rm a}$) shown in Fig. 11.

| $L_{\rm RC}$ | Redshift | b | a | $r_p$ |
|---|---|---|---|---|
| $\nu L_{1.3}$ | $1.5 < z < 3.5$ | $0.78 \pm 0.04$ | $-28.9 \pm 1.7$ | $0.82 \pm 0.04$ |
| $\nu L_{1.3}$ | $1.5 < z < 2$ | $0.78 \pm 0.08$ | $-29.1 \pm 3.3$ | $0.72 \pm 0.07$ |
| $\nu L_{1.3}$ | $2 < z < 3.5$ | $0.70 \pm 0.06$ | $-25.8 \pm 2.7$ | $0.78 \pm 0.07$ |
| $\nu L_3$ | $1.5 < z < 3.5$ | $0.82 \pm 0.03$ | $-30.5 \pm 1.5$ | $0.88 \pm 0.04$ |
| $\nu L_3$ | $1.5 < z < 2$ | $0.90 \pm 0.08$ | $-34.1 \pm 3.0$ | $0.83 \pm 0.06$ |
| $\nu L_3$ | $2 < z < 3.5$ | $0.83 \pm 0.06$ | $-28.5 \pm 2.5$ | $0.78 \pm 0.07$ |
| MRC | $1.5 < z < 3.5$ | $0.83 \pm 0.03$ | $-30.7 \pm 1.4$ | $0.90 \pm 0.03$ |
| MRC | $1.5 < z < 2$ | $0.91 \pm 0.07$ | $-34.5 \pm 3.0$ | $0.85 \pm 0.05$ |
| MRC | $2 < z < 3.5$ | $0.78 \pm 0.06$ | $-29.1 \pm 2.4$ | $0.85 \pm 0.05$ |

In the second method, we use the physically motivated SFR calibration relations in radio. Following Murphy

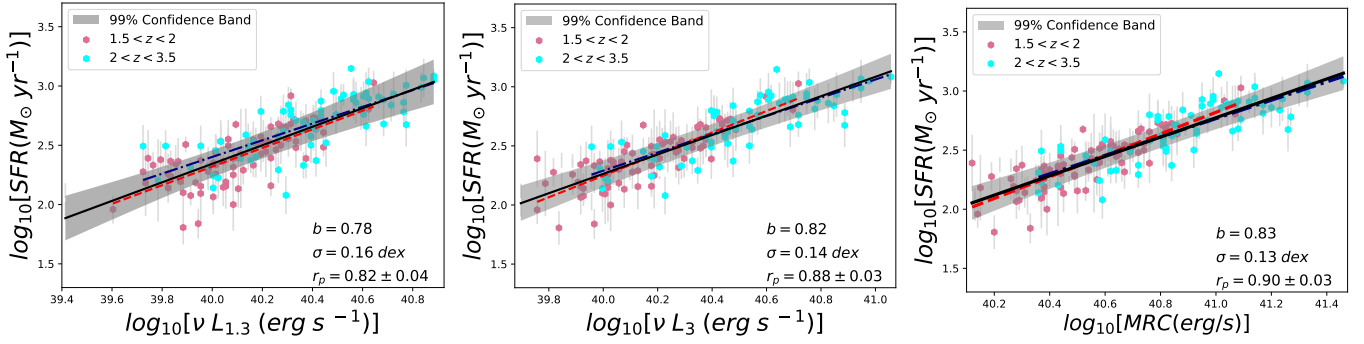

Figure 11: *Left-* The star formation rate (SFR) of the MIGHTEE-COSMOS SFGs located at $1.5 < z < 3.5$ against their rest-frame 1.3 GHz radio luminosity. Galaxies at $1.5 < z < 2$ are indicated in pink and at $2 < z < 4.5$ in blue. Lines show the linear regression fits to the entire points (black), at $1.5 < z < 2$ (red), and $2 < z < 3.5$ (blue). Its 99% confidence bound has been displayed in gray for the entire redshift intervals. *Middle-* Same as in the left panel but against the rest-frame 3 GHz radio luminosity. *Right-* Same as in the left panel, but against the integrated MRC luminosity.

Table 2: Correlation relations between TIR-based star formation rate, $SFR_{TIR}$, and that calibrated in radio, $SFR_{RC}$, given by $\log_{10}(SFR_{TIR}) = b \log_{10}(SFR_{RC}) + a$ (see Fig. 12).

| $SFR_{RC}$ | b | a | $r_p$ |
|---|---|---|---|
| $SFR_{1.3}$ | $0.75 \pm 0.04$ | $0.29 \pm 0.13$ | $0.80 \pm 0.04$ |
| $SFR_3$ | $0.78 \pm 0.04$ | $0.25 \pm 0.13$ | $0.81 \pm 0.04$ |
| $SFR_{MRC}$ | $1.00 \pm 0.04$ | $0.38 \pm 0.10$ | $0.90 \pm 0.03$ |

et al. (2011), the SFR is related to the RC luminosity at frequency $\nu$,

$$\left(\frac{SFR_\nu}{M_\odot \, yr^{-1}}\right) = 10^{-27} \left(\frac{L_\nu}{erg \, s^{-1} \, Hz^{-1}}\right) \times$$
$$\left[2.18 \left(\frac{T_e}{10^4 \, K}\right)^{0.45} \left(\frac{\nu}{GHz}\right)^{-0.1} + 15.1 \left(\frac{\nu}{GHz}\right)^{-\alpha_{nt}}\right]^{-1}, \quad (22)$$

with $T_e$ the electron temperature. Using the non-thermal spectral index $\alpha_{nt}$ and the 1.3 GHz luminosity $L_{1.3}$ and assuming $T_e = 10^4$ K, we obtained the $SFR_{1.3}$ and the $SFR_3$ for our MIGHTEE-COSMOS sample and compared it with $SFR_{TIR}$. A sub-linear correlation is found between the two SFR calibrators,

$$\left(\frac{SFR_{TIR}}{M_\odot \, yr^{-1}}\right) = 10^{(0.29 \pm 0.13)} \times \left(\frac{SFR_{1.3}}{M_\odot \, yr^{-1}}\right)^{(0.75 \pm 0.04)}, \quad (23)$$

$$\left(\frac{SFR_{TIR}}{M_\odot \, yr^{-1}}\right) = 10^{(0.25 \pm 0.13)} \times \left(\frac{SFR_3}{M_\odot \, yr^{-1}}\right)^{(0.78 \pm 0.04)}, \quad (24)$$

which agrees with the results of the first calibration method given by Eq.(19) and Eq.(20).

Tabatabaei et al. (2017) presented the following calibration relation using the integrated MRC luminosity for the KINGFISHER sample,

$$\left(\frac{SFR_{MRC}}{M_\odot \, yr^{-1}}\right) = 3.5 \times 10^{-31} \left(\frac{MRC}{erg \, s^{-1}}\right)^{(0.8 \pm 0.1)}. \quad (25)$$

This calibration is very similar to the one given by Eq.(21) for the high-z SFGs. Assuming that Eq.(25) also holds for the high-z SFGs, it is used to obtain the SFR for the MIGHTEE-COSMOS sample. Cross-correlating the resulting $SFR_{MRC}$ with the $SFR_{TIR}$ obtained from Eq.(21), the following linear correlation is obtained:

$$\left(\frac{SFR_{TIR}}{M_\odot \, yr^{-1}}\right) = 10^{(0.38 \pm 0.10)} \times \left(\frac{SFR_{MRC}}{M_\odot \, yr^{-1}}\right)^{(1.00 \pm 0.04)}, \quad (26)$$

with a Pearson correlation coefficient of $r_p \simeq 0.9$ (see also Fig. 12). This shows that the SFR calibration at $1.5 < z < 3.5$ agrees with that in the local universe within the uncertainties.

We note that Murphy et al. (2011) assumed a Kroupa IMF to obtain the expression given in Eq.(22). This IMF is similar to that of the Chabrier IMF used to obtain $SFR_{TIR}$ (Jin et al. 2018) for stars with masses $M_\star > 1 \, M_\odot$. Hence, the agreement obtained using methods 1 and 2 indicates that the stellar population in these high-z galaxies is dominated by stars more massive than the solar mass stars. As discussed by (Madau & Dickinson 2014) at young ages of star formation the luminosity is dominated by ultraviolet emission from massive stars and using the Kroupa vs Chabrier IMF to obtain the FUV conversion factor (converting the FUV luminosity to SFR) differs by less than 10% with age and metallicity.

The calibration relations obtained are sub-linear, meaning that the nonthermal emission is enhanced super-linearly with the SFR. Both the cosmic ray electron number and the magnetic field strength are increased due to massive star formation activities. Moreover, the energy of the cosmic ray electrons is relatively high and is similar to that of their injection energy (Sect. 6.1). The relatively flat synchrotron spectrum in the presence of a strong magnetic field is only possible if the magnetic field is turbulent, as discussed in previous studies (Tabatabaei et al. 2022, 2017). This is expected, with the small-scale turbulent dynamo acting as the main amplification mechanism of the magnetic field in these SFGs (see Sect. 6.2). The presence of strong turbulent magnetic fields has a consequence on the ISM energy balance: Those high-energy cosmic rays that scatter off the pitch angles of the turbulent field cause a pressure gradient in the ISM that leads to the creation of outflows and winds (Tabatabaei et al. 2018, 2022, called as 'nonthermal feedback').

Studying a sample of 178 galaxies from the MOSFIRE Deep Evolution Field (MOSDEF) survey at $1.4 < z < 2.6$, Duncan et al. (2020) de-reddened the H$\alpha$ fluxes to estimate the SFR and investigate the SFR–RC relation. They also found no systematic evolution with redshift, showing that the SFR–RC relation falls within a range of calibrations in the local universe.

Tabatabaei et al. (2017) showed that non-radio SFR calibrators such as the hybrid FUV+24$\mu$m and H$\alpha$+24$\mu$m agree better with SFR$_{\rm MRC}$ than with monochromatic RC calibrators. Similarly, we find that the correlation here is tighter with SFR$_{\rm MRC}$ than with SFR$_{1.3}$ and SFR$_3$. This is explained as the integrated MRC luminosity samples a wider range of age and energy of cosmic ray electrons and, hence, traces a more extended history of massive star formation compared to the monochromatic usage. Therefore, the use of the MRC luminosity as a robust SFR calibrator is emphasized at different redshifts through both calibration methods discussed in this section.

## 7. SUMMARY

We investigated the rest-frame mid-frequency radio SEDs for a sample of star-forming galaxies (SFGs) in the MIGHTEE-COSMOS field at redshift $1.5 < z < 3.5$ and estimated their equipartition magnetic field strength. Unlike frequent assumptions, we find that the shape of the radio SEDs is not fixed in the SFGs. This has an important consequence on the $k$-correction needed to obtain the radio luminosity of the high-z sources, study the IRRC, estimate the magnetic field strength, and calibrate the SFR over cosmic time. A summary of the most important findings are summarized:

- The mean nonthermal spectral index over the sample is $\alpha_{\rm nt} = 0.75 \pm 0.01$ (Sect. 3). The stacked $\alpha_{\rm nt}$ per redshift bins of 0.2 flattens with $z$ over $1.5 < z < 3.5$. It also flattens with the SFR per unit stellar mass in a similar way as of the normal star-forming galaxies at $z = 0$. The cosmic evolution of the SFR can explain the redshift evolution of the SEDs and that a younger and more energetic generation of cosmic ray electrons is found in galaxies with higher star formation activities (Sect. 6.1).

- With a sample mean of $\simeq 135\,\mu$G (Sect. 5), the equipartition magnetic field becomes stronger with $z$ obeying a power-law evolution of the form $B \simeq (1+z)^{0.7\pm0.1}$. The magnetic field is tightly correlated with the SFR following B $\simeq$ SFR$^{0.3}$, showing that it is mainly amplified by the act of the small-scale turbulent dynamo similar to that in nearby galaxies (Sect. 6.2).

- Investigating the IRRC shows that $q = 2.2 \pm 0.01$ is lower than that in the local universe (Sect. 6.3). This is linked to a super-linear enhancement of the synchrotron emission due to star formation activities, while the IR luminosity is linearly related to the SFR (Sect. 6.4). On the other hand, we find no systematic evolution of $q$ at the redshift range of $1.5 < z < 3.5$. Moreover, no correlation is found between $q$ and stellar mass.

- Integrating the rest frame SEDs over 1–10 GHz frequency range, the MRC luminosity is determined for the first time at high-z (Sect. 4). Calibrating the SFR using the RC luminosities and comparing them with the TIR-based measurement, we show that the MRC luminosity provides a more robust SFR tracer than monochromatic radio luminosities such as those at 1.3 GHz or 3 GHz (Sect. 6.4). We also present calibration relations between the MRC luminosity and a combination of the 1.3 GHz and 3 GHz luminosities.

As a concluding remark, the dominant nonthermal emission of this SFG population at mid-RC frequencies can be explained by a bright radio outflow component called nonthermal feedback because it is generated by the magnetic fields/cosmic ray pressures of highly star-forming regions. This picture explains not only the flat-tening of the synchrotron spectral index with star formation activity, but also the excess of the radio luminosity with respect to the TIR observed in the sample.

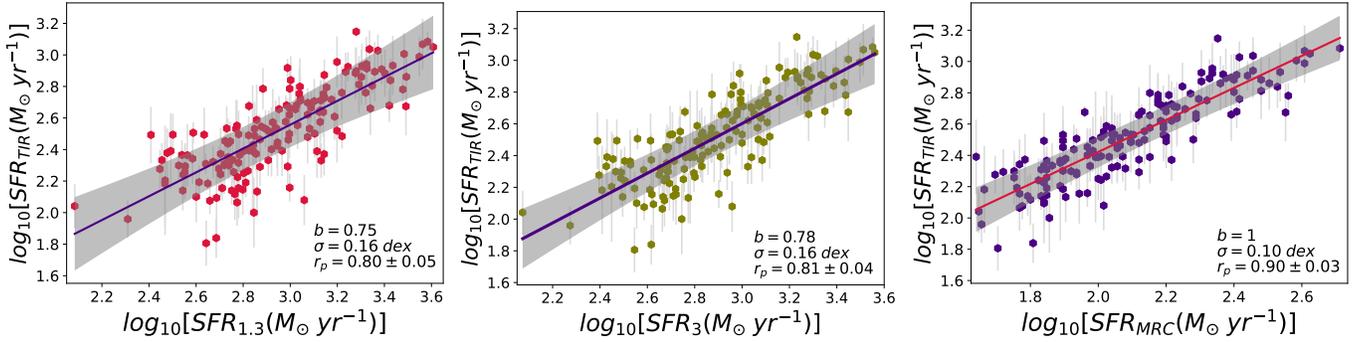

Figure 12: The SFR of the MIGHTEE-COSMOS SFGs at $1.5 < z < 3.5$ measured using the TIR luminosity against the SFRs calibrated in radio: *left-* 1.3 GHz, *middle-* 3 GHz, and *right-* MRC. Solid lines show the linear regression fits to the data in all panels. The gray shaded areas show their 99% confidence bounds.

Last but not least, the work presented in this paper will be benefited by the upcoming deep and multi-frequency SKA surveys, which will help to perform a more robust SED analysis in more complete samples.

We thank Catherine Hale and Ian Smail for their useful comments on the manuscript. FXA acknowledges the support from the National Natural Science Foundation of China (12303016) and the Natural Science Foundation of Jiangsu Province (BK20242115). We would also like to thank the anonymous referee for their constructive comments.

ACKNOWLEDGMENTS

# APPENDIX

## A. TABLES

Table A1. Data of MIGHTEE-COSMOS sample of star-forming galaxies with SNR> 1

| N | ID [1] | z [2] | $\log_{10}(M_{Star})$ $(M_\odot)$ | SFR $(M_\odot \mathrm{yr}^{-1})$ | $S_{1.3}^{(\mathrm{MeerKAT})}$ (mJy) | $S_{1.4}^{(\mathrm{VLA})}$ (mJy) | $S_{3}^{(VLA)}$ (mJy) | $S_{0.6}^{(\mathrm{GMRT})}$ (mJy) | $S_{0.3}^{(\mathrm{GMRT})}$ (mJy) |
|---|---|---|---|---|---|---|---|---|---|
| 1 | 286118 | 1.7 [i] | 10.79 | 291± 104 | 0.074± 0.004 | 0.086± 0.019 | 0.044± 0.003 | 0.224± 0.049 | 0.27± 0.122 |
| 2 | 292167 | 1.7 [i] | 10.59 | 479± 51 | 0.132± 0.01 | 0.142± 0.02 | 0.071± 0.004 | 0.156± 0.047 | 0.224± 0.164 |
| 3 | 307130 | 1.8 [i] | 10.91 | 324± 118 | 0.062± 0.005 | 0.059± 0.019 | 0.033± 0.003 | 0.094± 0.044 | 0.307± 0.111 |
| 4 | 322176 | 1.8 [i] | 10.91 | 283± 82 | 0.037± 0.007 | 0.054± 0.019 | 0.02± 0.003 | 0.063± 0.045 | 0.239± 0.138 |
| 5 | 338500 | 2 [i] | 11.1 | 811± 176 | 0.118± 0.012 | 0.116± 0.02 | 0.057± 0.004 | 0.194± 0.043 | 0.291± 0.09 |
| 6 | 345185 | 3.1 [i] | 11.43 | 568± 212 | 0.049± 0.003 | 0.056± 0.017 | 0.028± 0.003 | 0.101± 0.044 | 0.131± 0.11 |
| 7 | 350085 | 1.6 [ii] | 11.1 | 119± 23 | 0.045± 0.006 | 0.038± 0.017 | 0.015± 0.003 | 0.086± 0.043 | 0.1± 0.1 |
| 8 | 356308 | 2 [i] | 11.09 | 144± 29 | 0.033± 0.005 | 0.023± 0.017 | 0.013± 0.002 | 0.096± 0.041 | 0.199± 0.09 |
| 9 | 363438 | 2.1 [i] | 11.51 | 848± 281 | 0.048± 0.007 | 0.051± 0.015 | 0.02± 0.003 | 0.072± 0.039 | 0.14± 0.14 |
| 10 | 375061 | 1.9 [i] | 10.98 | 539± 103 | 0.081± 0.008 | 0.053± 0.016 | 0.032± 0.003 | 0.113± 0.041 | 0.225± 0.133 |
| 11 | 382482 | 1.7 [ii] | 11.03 | 181± 38 | 0.043± 0.005 | 0.036± 0.017 | 0.018± 0.003 | 0.056± 0.043 | 0.24± 0.153 |
| 12 | 388558 | 2.9 [ii] | 10.8 | 478± 114 | 0.061± 0.007 | 0.051± 0.014 | 0.028± 0.003 | 0.093± 0.04 | 0.258± 0.106 |
| 13 | 396371 | 2 [i] | 11.19 | 638± 73 | 0.178± 0.01 | 0.166± 0.016 | 0.087± 0.005 | 0.249± 0.04 | 0.582± 0.106 |
| 14 | 398977 | 1.6 [i] | 10.98 | 246± 135 | 0.029± 0.009 | 0.025± 0.014 | 0.012± 0.002 | 0.045± 0.039 | 0.137± 0.085 |
| 15 | 406998 | 1.8 [i] | 10.29 | 189± 53 | 0.024± 0.004 | 0.027± 0.016 | 0.014± 0.002 | 0.04± 0.04 | 0.145± 0.089 |
| 16 | 410945 | 1.9 [i] | 11.76 | 527± 33 | 0.078± 0.006 | 0.074± 0.015 | 0.044± 0.003 | 0.143± 0.038 | 0.264± 0.105 |
| 17 | 411577 | 1.8 [i] | 10.66 | 100± 51 | 0.033± 0.004 | 0.025± 0.017 | 0.013± 0.002 | 0.046± 0.043 | 0.23± 0.093 |
| 18 | 415269 | 1.6 [i] | 10.48 | 153± 13 | 0.048± 0.009 | 0.05± 0.016 | 0.029± 0.003 | 0.094± 0.039 | 0.113± 0.104 |
| 19 | 427827 | 2.4 [i] | 11.38 | 765± 291 | 0.073± 0.005 | 0.068± 0.017 | 0.04± 0.003 | 0.079± 0.041 | 0.153± 0.097 |
| 20 | 439437 | 3.4 [i] | 10.44 | 820± 239 | 0.056± 0.008 | 0.025± 0.016 | 0.024± 0.003 | 0.061± 0.039 | 0.08± 0.08 |
| 21 | 441820 | 3 [i] | 10.82 | 1168± 436 | 0.041± 0.007 | 0.05± 0.017 | 0.024± 0.003 | 0.075± 0.041 | 0.15± 0.1 |
| 22 | 442090 | 1.6 [i] | 11.24 | 321± 49 | 0.061± 0.018 | 0.068± 0.017 | 0.043± 0.003 | 0.079± 0.043 | 0.124± 0.108 |
| 23 | 449963 | 2.6 [i] | 10.99 | 311± 128 | 0.016± 0.004 | 0.019± 0.017 | 0.014± 0.002 | 0.064± 0.037 | 0.093± 0.08 |
| 24 | 452505 | 1.9 [i] | 11.24 | 231± 105 | 0.028± 0.011 | 0.012± 0.015 | 0.02± 0.003 | 0.06± 0.036 | 0.117± 0.072 |
| 25 | 462193 | 1.9 [ii] | 10.78 | 328± 100 | 0.05± 0.011 | 0.048± 0.016 | 0.024± 0.003 | 0.045± 0.037 | 0.327± 0.102 |
| 26 | 469122 | 3.2 [i] | 10.76 | 340± 143 | 0.029± 0.005 | 0.01± 0.008 | 0.014± 0.003 | 0.101± 0.047 | 0.1± 0.1 |
| 27 | 470051 | 1.9 [i] | 10.44 | 342± 110 | 0.051± 0.005 | 0.059± 0.025 | 0.035± 0.003 | 0.061± 0.054 | 0.12± 0.1 |
| 28 | 474662 | 2 [i] | 10.83 | 459± 183 | 0.029± 0.004 | 0.022± 0.017 | 0.015± 0.002 | 0.041± 0.035 | 0.119± 0.074 |
| 29 | 476581 | 3.2 [i] | 10.84 | 576± 221 | 0.068± 0.005 | 0.072± 0.017 | 0.032± 0.003 | 0.048± 0.039 | 0.186± 0.124 |
| 30 | 477095 | 1.8 [i] | 10.64 | 316± 114 | 0.088± 0.005 | 0.075± 0.016 | 0.041± 0.003 | 0.07± 0.041 | 0.156± 0.087 |
| 31 | 478096 | 1.8 [i] | 10.7 | 236± 89 | 0.031± 0.005 | 0.011± 0.016 | 0.013± 0.002 | 0.084± 0.037 | 0.138± 0.072 |
| 32 | 486997 | 1.8 [i] | 10.65 | 434± 158 | 0.089± 0.007 | 0.09± 0.016 | 0.042± 0.003 | 0.092± 0.037 | 0.28± 0.099 |
| 33 | 488383 | 1.8 [i] | 10.04 | 215± 54 | 0.033± 0.005 | 0.021± 0.016 | 0.022± 0.003 | 0.033± 0.03 | 0.11± 0.11 |
| 34 | 495704 | 2.1 [ii] | 11.28 | 902± 337 | 0.115± 0.01 | 0.114± 0.02 | 0.055± 0.004 | 0.127± 0.035 | 0.208± 0.069 |

[1] Identifie from COSMOS Super-deblended Photometry Catalog (Jin et al. 2018)

[2] Photometric Redshift from COSMOS Catalogues (Weaver et al. 2022; Laigle et al. 2016)

[i] Photometric Redshift from COSMOS 2020 Catalogue (Weaver et al. 2022)

[ii] Photometric Redshift from COSMOS 2015 Catalogue (Laigle et al. 2016)

| 36 | 501111 | 2.4 [i]  | 11    | 874± 353  | 0.06± 0.007  | 0.07± 0.016  | 0.04± 0.003  | 0.15± 0.04   | 0.163± 0.119 |
|----|--------|----------|-------|-----------|--------------|--------------|--------------|--------------|--------------|
| 37 | 502532 | 1.9 [i]  | 10.21 | 181± 72   | 0.021± 0.003 | 0.011± 0.016 | 0.012± 0.002 | 0.045± 0.036 | 0.169± 0.083 |
| 38 | 502934 | 1.7 [i]  | 11.14 | 195± 54   | 0.04± 0.004  | 0.045± 0.018 | 0.02± 0.003  | 0.136± 0.04  | 0.228± 0.093 |
| 39 | 506989 | 1.6 [i]  | 10.73 | 69± 20    | 0.032± 0.005 | 0.03± 0.02   | 0.016± 0.003 | 0.071± 0.041 | 0.224± 0.12  |
| 40 | 516049 | 1.6 [i]  | 10.71 | 64± 21    | 0.04± 0.008  | 0.042± 0.016 | 0.012± 0.002 | 0.073± 0.04  | 0.098± 0.093 |
| 41 | 516610 | 1.6 [ii] | 10.09 | 91± 25    | 0.027± 0.004 | 0.032± 0.02  | 0.017± 0.003 | 0.039± 0.046 | 0.111± 0.098 |
| 42 | 516810 | 1.6 [i]  | 10.63 | 245± 75   | 0.071± 0.004 | 0.074± 0.016 | 0.04± 0.003  | 0.131± 0.039 | 0.223± 0.102 |
| 43 | 525980 | 2.1 [ii] | 10.92 | 312± 96   | 0.028± 0.003 | 0.03± 0.015  | 0.021± 0.003 | 0.044± 0.045 | 0.149± 0.099 |
| 44 | 526965 | 1.6 [ii] | 11.34 | 143± 26   | 0.039± 0.005 | 0.021± 0.015 | 0.014± 0.002 | 0.048± 0.035 | 0.2± 0.077   |
| 45 | 529975 | 2.5 [ii] | 10.44 | 196± 50   | 0.032± 0.005 | 0.028± 0.015 | 0.019± 0.003 | 0.109± 0.039 | 0.091± 0.124 |
| 46 | 533924 | 3.1 [i]  | 11.47 | 576± 103  | 0.059± 0.008 | 0.045± 0.016 | 0.03± 0.003  | 0.058± 0.037 | 0.158± 0.102 |
| 47 | 540343 | 2.9 [i]  | 9.69  | 138± 28   | 0.042± 0.005 | 0.037± 0.018 | 0.017± 0.002 | 0.083± 0.041 | 0.13± 0.102  |
| 48 | 547102 | 1.6 [ii] | 10.77 | 138± 40   | 0.043± 0.005 | 0.042± 0.017 | 0.023± 0.003 | 0.053± 0.034 | 0.144± 0.094 |
| 49 | 556782 | 1.9 [i]  | 10.89 | 232± 94   | 0.036± 0.003 | 0.045± 0.015 | 0.014± 0.002 | 0.03± 0.03   | 0.097± 0.064 |
| 50 | 559532 | 1.7 [i]  | 10.82 | 220± 81   | 0.037± 0.006 | 0.032± 0.016 | 0.016± 0.002 | 0.053± 0.035 | 0.074± 0.087 |
| 51 | 561437 | 2.4 [i]  | 11.07 | 746± 167  | 0.065± 0.014 | 0.039± 0.015 | 0.037± 0.003 | 0.071± 0.04  | 0.251± 0.076 |
| 52 | 564732 | 1.8 [i]  | 10.7  | 314± 108  | 0.048± 0.003 | 0.049± 0.015 | 0.027± 0.003 | 0.064± 0.035 | 0.142± 0.078 |
| 53 | 565719 | 1.9 [i]  | 11.11 | 344± 140  | 0.043± 0.006 | 0.055± 0.015 | 0.019± 0.003 | 0.047± 0.035 | 0.126± 0.08  |
| 54 | 565792 | 1.9 [i]  | 10.81 | 406± 137  | 0.074± 0.007 | 0.069± 0.018 | 0.024± 0.003 | 0.126± 0.035 | 0.134± 0.073 |
| 55 | 569615 | 2.8 [i]  | 11.46 | 623± 31   | 0.067± 0.007 | 0.064± 0.016 | 0.035± 0.003 | 0.108± 0.038 | 0.187± 0.101 |
| 56 | 575519 | 1.9 [i]  | 10.75 | 232± 49   | 0.028± 0.006 | 0.033± 0.016 | 0.019± 0.002 | 0.054± 0.034 | 0.107± 0.101 |
| 57 | 578926 | 2.4 [i]  | 10.96 | 724± 295  | 0.093± 0.006 | 0.064± 0.016 | 0.04± 0.003  | 0.095± 0.036 | 0.23± 0.086  |
| 58 | 579041 | 1.8 [i]  | 10.46 | 178± 55   | 0.06± 0.007  | 0.052± 0.017 | 0.031± 0.003 | 0.04± 0.039  | 0.227± 0.093 |
| 59 | 580926 | 1.6 [ii] | 10.57 | 152± 97   | 0.027± 0.004 | 0.051± 0.014 | 0.013± 0.002 | 0.046± 0.032 | 0.1± 0.1     |
| 60 | 584422 | 2.8 [i]  | 10.77 | 856± 371  | 0.072± 0.005 | 0.065± 0.018 | 0.041± 0.003 | 0.121± 0.04  | 0.149± 0.08  |
| 61 | 590368 | 2.4 [i]  | 11.21 | 512± 188  | 0.03± 0.012  | 0.022± 0.016 | 0.022± 0.003 | 0.083± 0.035 | 0.107± 0.085 |
| 62 | 595018 | 1.7 [i]  | 10.46 | 190± 62   | 0.063± 0.01  | 0.05± 0.015  | 0.032± 0.003 | 0.064± 0.036 | 0.174± 0.085 |
| 63 | 595222 | 1.6 [ii] | 10.62 | 174± 56   | 0.047± 0.003 | 0.051± 0.016 | 0.026± 0.003 | 0.014± 0.003 | 0.133± 0.078 |
| 65 | 612581 | 1.8 [i]  | 10.55 | 274± 100  | 0.052± 0.006 | 0.027± 0.019 | 0.025± 0.003 | 0.054± 0.037 | 0.127± 0.092 |
| 66 | 616280 | 3 [ii]   | 11.4  | 1214± 173 | 0.159± 0.011 | 0.153± 0.02  | 0.093± 0.005 | 0.168± 0.044 | 0.255± 0.1   |
| 67 | 623091 | 2.2 [ii] | 11.24 | 515± 146  | 0.063± 0.006 | 0.06± 0.018  | 0.035± 0.003 | 0.118± 0.043 | 0.286± 0.084 |
| 68 | 625223 | 1.8 [i]  | 10.29 | 239± 85   | 0.041± 0.003 | 0.042± 0.016 | 0.031± 0.003 | 0.061± 0.035 | 0.1± 0.1     |
| 69 | 627356 | 2.2 [i]  | 11.49 | 973± 280  | 0.11± 0.004  | 0.09± 0.016  | 0.06± 0.004  | 0.129± 0.036 | 0.292± 0.076 |
| 70 | 628033 | 1.8 [i]  | 11.13 | 509± 165  | 0.058± 0.006 | 0.05± 0.016  | 0.028± 0.003 | 0.082± 0.035 | 0.146± 0.095 |
| 71 | 631682 | 3.3 [i]  | 10.84 | 1087± 297 | 0.065± 0.007 | 0.044± 0.016 | 0.034± 0.003 | 0.041± 0.036 | 0.236± 0.103 |
| 72 | 632758 | 1.8 [i]  | 11.46 | 341± 68   | 0.072± 0.004 | 0.044± 0.018 | 0.021± 0.003 | 0.122± 0.036 | 0.082± 0.067 |
| 73 | 634466 | 3.3 [ii] | 10.49 | 1404± 80  | 0.04± 0.013  | 0.027± 0.018 | 0.019± 0.003 | 0.077± 0.033 | 0.1± 0.1     |
| 74 | 634999 | 1.8 [i]  | 10.88 | 155± 33   | 0.061± 0.005 | 0.025± 0.016 | 0.02± 0.003  | 0.102± 0.046 | 0.146± 0.126 |
| 75 | 635353 | 2.6 [i]  | 10.81 | 331± 154  | 0.039± 0.007 | 0.029± 0.016 | 0.017± 0.003 | 0.053± 0.039 | 0.203± 0.074 |
| 76 | 635836 | 1.7 [ii] | 11.48 | 269± 86   | 0.062± 0.005 | 0.053± 0.017 | 0.03± 0.003  | 0.09± 0.035  | 0.212± 0.096 |
| 77 | 638136 | 1.5 [i]  | 10.61 | 197± 66   | 0.031± 0.006 | 0.038± 0.016 | 0.016± 0.002 | 0.072± 0.036 | 0.114± 0.077 |
| 78 | 638642 | 2 [i]    | 11.1  | 284± 73   | 0.058± 0.01  | 0.06± 0.018  | 0.033± 0.003 | 0.064± 0.034 | 0.184± 0.092 |
| 79 | 644839 | 1.6 [i]  | 11    | 216± 71   | 0.051± 0.008 | 0.015± 0.016 | 0.019± 0.002 | 0.072± 0.038 | 0.153± 0.117 |
| 80 | 645724 | 2.3 [i]  | 11.36 | 726± 281  | 0.074± 0.006 | 0.082± 0.018 | 0.036± 0.003 | 0.122± 0.042 | 0.219± 0.079 |
| 81 | 647970 | 1.9 [i]  | 10.76 | 140± 51   | 0.029± 0.011 | 0.019± 0.018 | 0.013± 0.002 | 0.068± 0.041 | 0.191± 0.103 |
| 82 | 651173 | 1.8 [i]  | 11.2  | 533± 181  | 0.11± 0.016  | 0.106± 0.017 | 0.05± 0.003  | 0.166± 0.034 | 0.368± 0.094 |
| 83 | 651584 | 2.3 [i]  | 10.88 | 564± 143  | 0.084± 0.013 | 0.081± 0.016 | 0.047± 0.003 | 0.111± 0.046 | 0.11± 0.1    |
| 84 | 657953 | 1.5 [i]  | 10.4  | 184± 4    | 0.085± 0.008 | 0.082± 0.017 | 0.05± 0.004  | 0.141± 0.033 | 0.239± 0.071 |

| 85 | 662169 | 2.2 [i] | 10.64 | 214± 54 | 0.041± 0.011 | 0.057± 0.018 | 0.016± 0.002 | 0.098± 0.04 | 0.298± 0.088 |
|---|---|---|---|---|---|---|---|---|---|
| 86 | 662536 | 2.6 [i] | 10.9 | 780± 218 | 0.054± 0.007 | 0.048± 0.017 | 0.031± 0.003 | 0.149± 0.04 | 0.123± 0.11 |
| 87 | 678384 | 1.8 [i] | 11.29 | 474± 39 | 0.064± 0.015 | 0.065± 0.02 | 0.022± 0.003 | 0.082± 0.042 | 0.091± 0.084 |
| 88 | 680956 | 2.7 [ii] | 11.34 | 782± 339 | 0.06± 0.006 | 0.07± 0.016 | 0.029± 0.003 | 0.087± 0.039 | 0.28± 0.089 |
| 89 | 690848 | 2 [i] | 10.48 | 387± 199 | 0.039± 0.007 | 0.021± 0.018 | 0.019± 0.003 | 0.045± 0.037 | 0.179± 0.148 |
| 90 | 692002 | 2.9 [i] | 11.04 | 604± 112 | 0.056± 0.005 | 0.062± 0.017 | 0.03± 0.003 | 0.063± 0.035 | 0.1± 0.1 |
| 91 | 696690 | 2.7 [i] | 10.42 | 120± 44 | 0.045± 0.005 | 0.02± 0.019 | 0.014± 0.003 | 0.085± 0.044 | 0.213± 0.094 |
| 92 | 705179 | 1.7 | 10.9 | 340± 93 | 0.062± 0.005 | 0.062± 0.017 | 0.027± 0.003 | 0.114± 0.039 | 0.079± 0.077 |
| 93 | 705616 | 1.9 [i] | 10.97 | 279± 137 | 0.04± 0.005 | 0.048± 0.015 | 0.02± 0.003 | 0.093± 0.041 | 0.134± 0.08 |
| 94 | 707850 | 3.2 [i] | 10.42 | 309± 86 | 0.028± 0.004 | 0.011± 0.016 | 0.021± 0.003 | 0.01± 0.038 | 0.1± 0.1 |
| 95 | 708131 | 2 [i] | 10.96 | 789± 268 | 0.145± 0.01 | 0.128± 0.018 | 0.078± 0.004 | 0.248± 0.037 | 0.312± 0.102 |
| 96 | 710739 | 1.9 [i] | 9.68 | 136± 50 | 0.037± 0.004 | 0.019± 0.017 | 0.016± 0.002 | 0.065± 0.035 | 0.196± 0.099 |
| 97 | 717183 | 1.9 [i] | 10.8 | 266± 90 | 0.039± 0.006 | 0.037± 0.016 | 0.017± 0.003 | 0.114± 0.035 | 0.156± 0.087 |
| 98 | 718344 | 2.9 [i] | 10.17 | 418± 162 | 0.031± 0.005 | 0.023± 0.016 | 0.018± 0.002 | 0.094± 0.039 | 0.125± 0.089 |
| 99 | 727184 | 2 [i] | 9.97 | 245± 236 | 0.042± 0.005 | 0.032± 0.021 | 0.022± 0.003 | 0.083± 0.051 | 0.159± 0.106 |
| 100 | 729926 | 1.8 [i] | 10.52 | 124± 73 | 0.028± 0.004 | 0.019± 0.015 | 0.013± 0.002 | 0.043± 0.039 | 0.093± 0.088 |
| 101 | 730319 | 2.9 [ii] | 11.2 | 649± 270 | 0.081± 0.005 | 0.083± 0.016 | 0.053± 0.004 | 0.085± 0.036 | 0.209± 0.074 |
| 102 | 732468 | 2.7 [ii] | 10.67 | 307± 139 | 0.021± 0.008 | 0.047± 0.016 | 0.02± 0.002 | 0.057± 0.042 | 0.162± 0.091 |
| 104 | 748925 | 3.2 [i] | 10.23 | 168± 67 | 0.027± 0.004 | 0.032± 0.015 | 0.011± 0.002 | 0.054± 0.04 | 0.123± 0.105 |
| 105 | 749578 | 1.7 [ii] | 11.16 | 521± 33 | 0.12± 0.009 | 0.12± 0.017 | 0.048± 0.003 | 0.193± 0.043 | 0.366± 0.099 |
| 106 | 751335 | 1.6 [i] | 10.73 | 479± 52 | 0.121± 0.018 | 0.108± 0.019 | 0.056± 0.004 | 0.139± 0.039 | 0.533± 0.295 |
| 107 | 757799 | 1.6 [i] | 10.92 | 126± 47 | 0.047± 0.005 | 0.054± 0.017 | 0.026± 0.003 | 0.069± 0.038 | 0.1± 0.1 |
| 108 | 762527 | 2.1 [i] | 11.27 | 204± 31 | 0.052± 0.006 | 0.027± 0.018 | 0.023± 0.003 | 0.035± 0.041 | 0.108± 0.096 |
| 109 | 770288 | 2.1 [i] | 11.15 | 210± 50 | 0.052± 0.005 | 0.063± 0.015 | 0.038± 0.003 | 0.045± 0.043 | 0.177± 0.127 |
| 110 | 772906 | 2 [i] | 10.37 | 340± 133 | 0.054± 0.003 | 0.056± 0.017 | 0.039± 0.003 | 0.007± 0.037 | 0.115± 0.072 |
| 112 | 799864 | 2.4 [i] | 10.65 | 235± 73 | 0.065± 0.005 | 0.064± 0.016 | 0.034± 0.003 | 0.138± 0.047 | 0.243± 0.115 |
| 113 | 801243 | 1.8 [i] | 11.22 | 248± 61 | 0.056± 0.004 | 0.058± 0.019 | 0.03± 0.003 | 0.088± 0.044 | 0.165± 0.095 |
| 114 | 803543 | 2.6 [ii] | 10.98 | 376± 168 | 0.04± 0.009 | 0.041± 0.016 | 0.021± 0.003 | 0.079± 0.043 | 0.179± 0.1 |
| 115 | 808735 | 1.5 [ii] | 9.87 | 110± 35 | 0.042± 0.039 | 0.026± 0.017 | 0.022± 0.003 | 0.022± 0.04 | 0.03± 0.01 |
| 116 | 819326 | 2.6 [ii] | 10.53 | 435± 150 | 0.062± 0.015 | 0.076± 0.018 | 0.031± 0.003 | 0.052± 0.041 | 0.082± 0.08 |
| 117 | 827300 | 1.5 [ii] | 11.09 | 199± 41 | 0.073± 0.007 | 0.065± 0.017 | 0.034± 0.003 | 0.119± 0.042 | 0.123± 0.082 |
| 118 | 837455 | 1.8 [i] | 10.74 | 356± 132 | 0.067± 0.014 | 0.068± 0.017 | 0.034± 0.003 | 0.06± 0.039 | 0.12± 0.107 |
| 119 | 838572 | 2.7 [i] | 10.96 | 610± 186 | 0.044± 0.005 | 0.041± 0.018 | 0.031± 0.003 | 0.047± 0.04 | 0.096± 0.084 |
| 120 | 842239 | 1.7 [i] | 10.53 | 204± 52 | 0.04± 0.009 | 0.061± 0.017 | 0.021± 0.003 | 0.076± 0.045 | 0.295± 0.163 |
| 121 | 842950 | 2.5 [i] | 10.66 | 629± 260 | 0.033± 0.004 | 0.026± 0.019 | 0.018± 0.002 | 0.067± 0.042 | 0.217± 0.103 |
| 122 | 849028 | 2.9 [i] | 10.77 | 1014± 311 | 0.113± 0.011 | 0.111± 0.016 | 0.056± 0.004 | 0.106± 0.042 | 0.127± 0.09 |
| 123 | 851363 | 2 [ii] | 10.84 | 458± 178 | 0.141± 0.005 | 0.127± 0.017 | 0.067± 0.004 | 0.173± 0.048 | 0.344± 0.105 |
| 124 | 852900 | 2.4 [i] | 10.92 | 160± 128 | 0.018± 0.007 | 0.044± 0.017 | 0.019± 0.003 | 0.078± 0.047 | 0.086± 0.102 |
| 125 | 863440 | 1.6 [i] | 10.58 | 116± 39 | 0.038± 0.004 | 0.03± 0.015 | 0.014± 0.002 | 0.058± 0.043 | 0.27± 0.11 |
| 126 | 871000 | 1.7 [i] | 11.21 | 305± 55 | 0.053± 0.005 | 0.039± 0.019 | 0.013± 0.003 | 0.103± 0.044 | 0.223± 0.121 |
| 127 | 871582 | 1.5 [i] | 11.28 | 131± 49 | 0.031± 0.014 | 0.029± 0.017 | 0.019± 0.003 | 0.169± 0.045 | 0.256± 0.118 |
| 129 | 902320 | 1.7 [i] | 10.98 | 1064± 250 | 0.227± 0.043 | 0.186± 0.024 | 0.087± 0.003 | 0.308± 0.048 | 0.384± 0.198 |
| 130 | 10033174 | 2.8 [ii] | 9.99 | 840± 181 | 0.049± 0.007 | 0.046± 0.018 | 0.031± 0.003 | 0.097± 0.04 | 0.328± 0.097 |
| 131 | 10035127 | 2 [ii] | 11.12 | 462± 107 | 0.117± 0.011 | 0.117± 0.017 | 0.063± 0.004 | 0.126± 0.039 | 0.122± 0.113 |
| 132 | 10040313 | 1.8 [ii] | 9.97 | 247± 91 | 0.073± 0.005 | 0.079± 0.017 | 0.032± 0.003 | 0.05± 0.039 | 0.189± 0.092 |
| 133 | 10070338 | 1.7 [ii] | 10.2 | 207± 68 | 0.094± 0.007 | 0.09± 0.018 | 0.052± 0.004 | 0.001± 0.039 | 0.104± 0.084 |
| 134 | 10078195 | 1.6 [ii] | 10.47 | 351± 21 | 0.091± 0.005 | 0.088± 0.019 | 0.05± 0.003 | 0.117± 0.044 | 0.177± 0.106 |
| 135 | 10080288 | 2.3 [ii] | 11.25 | 445± 197 | 0.064± 0.009 | 0.072± 0.017 | 0.031± 0.003 | 0.059± 0.042 | 0.215± 0.101 |

| 136 | 10085717 | 1.7 ii | 10.77 | 200± 84 | 0.044± 0.006 | 0.052± 0.019 | 0.024± 0.003 | 0.08± 0.04 | 0.155± 0.112 |
|---|---|---|---|---|---|---|---|---|---|
| 137 | 10090559 | 1.5 ii | 10.62 | 199± 41 | 0.039± 0.005 | 0.036± 0.017 | 0.021± 0.003 | 0.055± 0.04 | 0.166± 0.092 |
| 138 | 10090785 | 2 ii | 11.15 | 377± 145 | 0.064± 0.006 | 0.058± 0.016 | 0.039± 0.003 | 0.189± 0.04 | 0.14± 0.109 |
| 140 | 10093049 | 1.6 ii | 10.81 | 225± 69 | 0.062± 0.007 | 0.072± 0.017 | 0.027± 0.003 | 0.091± 0.042 | 0.115± 0.079 |
| 141 | 10093193 | 2.3 ii | 11.65 | 765± 262 | 0.094± 0.013 | 0.12± 0.017 | 0.069± 0.004 | 0.273± 0.048 | 0.444± 0.137 |
| 142 | 10094402 | 1.8 ii | 10.62 | 245± 95 | 0.106± 0.033 | 0.099± 0.017 | 0.036± 0.003 | 0.069± 0.048 | 0.407± 0.114 |
| 143 | 10094777 | 1.6 ii | 10.67 | 290± 95 | 0.089± 0.008 | 0.096± 0.017 | 0.038± 0.003 | 0.122± 0.04 | 0.321± 0.112 |
| 144 | 10095574 | 1.8 ii | 10.15 | 340± 104 | 0.079± 0.006 | 0.061± 0.017 | 0.04± 0.003 | 0.09± 0.044 | 0.104± 0.118 |
| 145 | 10103325 | 2.3 ii | 10.87 | 472± 200 | 0.177± 0.017 | 0.171± 0.02 | 0.085± 0.005 | 0.226± 0.037 | 0.434± 0.096 |
| 146 | 10123324 | 2.3 ii | 10.82 | 317± 111 | 0.077± 0.006 | 0.064± 0.016 | 0.045± 0.003 | 0.137± 0.035 | 0.138± 0.085 |
| 147 | 10124598 | 2.1 ii | 10.41 | 497± 86 | 0.091± 0.008 | 0.06± 0.016 | 0.047± 0.003 | 0.08± 0.035 | 0.22± 0.127 |
| 149 | 10135986 | 2 ii | 10.41 | 252± 100 | 0.039± 0.005 | 0.024± 0.015 | 0.018± 0.002 | 0.089± 0.036 | 0.202± 0.089 |
| 150 | 10137050 | 2 ii | 10.25 | 312± 123 | 0.047± 0.004 | 0.047± 0.016 | 0.037± 0.003 | 0.074± 0.038 | 0.139± 0.099 |
| 153 | 10149158 | 2.5 ii | 11.12 | 1121± 55 | 0.177± 0.021 | 0.173± 0.022 | 0.077± 0.005 | 0.219± 0.044 | 0.433± 0.098 |
| 154 | 10157709 | 1.9 ii | 10.44 | 302± 117 | 0.074± 0.007 | 0.049± 0.019 | 0.029± 0.003 | 0.052± 0.043 | 0.301± 0.121 |
| 155 | 10176688 | 1.5 ii | 11.04 | 246± 65 | 0.098± 0.006 | 0.107± 0.017 | 0.046± 0.003 | 0.124± 0.042 | 0.011± 0.09 |
| 156 | 10205913 | 2.2 ii | 10.82 | 236± 60 | 0.05± 0.007 | 0.054± 0.015 | 0.034± 0.003 | 0.087± 0.04 | 0.127± 0.094 |
| 157 | 10209718 | 1.5 ii | 10.4 | 143± 63 | 0.082± 0.008 | 0.06± 0.017 | 0.036± 0.003 | 0.121± 0.041 | 0.167± 0.082 |
| 158 | 10211581 | 3.2 ii | 10.92 | 611± 199 | 0.024± 0.005 | 0.032± 0.014 | 0.016± 0.003 | 0.046± 0.042 | 0.148± 0.121 |
| 159 | 10219023 | 2.6 ii | 10.84 | 690± 216 | 0.117± 0.006 | 0.098± 0.021 | 0.061± 0.004 | 0.087± 0.043 | 0.213± 0.101 |
| 160 | 10219338 | 1.8 ii | 10.74 | 156± 38 | 0.027± 0.006 | 0.027± 0.019 | 0.012± 0.002 | 0.058± 0.043 | 0.086± 0.094 |

Table A2. SED parameters $\alpha_{nt}$, $F_{th(1.3)}$, rest-frame radio luminosities $L_{1.3}$, $L_3$, MRC.

| N | ID [1] | z [2] | $\alpha_{nt}$ | $F_{th}$ | $\log_{10}(\mathrm{MRC\,(erg/s)})$ | $\log_{10}(L_{1.3}\,(\mathrm{erg/s\,Hz}))$ | $\log_{10}(L_3\,(\mathrm{erg/s\,Hz}))$ |
|---|---|---|---|---|---|---|---|
| 1 | 286118 | 1.7 | $0.70^{0.10}_{0.08}$ | – | 40.64 | $31.04^{0.07}_{0.06}$ | $30.79^{0.07}_{0.06}$ |
| 2 | 292167 | 1.7 | $0.65^{0.07}_{0.07}$ | – | 40.82 | $31.20^{0.06}_{0.06}$ | $30.96^{0.06}_{0.06}$ |
| 3 | 307130 | 1.8 | $0.79^{0.10}_{0.09}$ | – | 40.65 | $31.08^{0.07}_{0.06}$ | $30.80^{0.07}_{0.06}$ |
| 4 | 322176 | 1.8 | $0.84^{0.15}_{0.18}$ | – | 40.44 | $30.90^{0.11}_{0.12}$ | $30.00^{0.11}_{0.12}$ |
| 5 | 338500 | 2.0 | $0.77^{0.07}_{0.08}$ | – | 41.12 | $31.55^{0.05}_{0.05}$ | $31.27^{0.05}_{0.05}$ |
| 6 | 345185 | 3.1 | $0.77^{0.21}_{0.25}$ | $0.08^{0.24}_{0.34}$ | 40.79 | $31.20^{1.49}_{2.23}$ | $30.93^{1.49}_{2.23}$ |
| 7 | 350085 | 1.6 | $0.98^{0.14}_{0.15}$ | – | 40.29 | $30.80^{0.08}_{0.09}$ | $30.45^{0.08}_{0.09}$ |
| 8 | 356308 | 2.0 | $1.18^{0.15}_{0.15}$ | – | 40.53 | $31.12^{0.08}_{0.08}$ | $30.70^{0.08}_{0.08}$ |
| 9 | 363438 | 2.1 | $0.94^{0.13}_{0.14}$ | – | 40.95 | $31.45^{0.08}_{0.09}$ | $31.10^{0.08}_{0.09}$ |
| 10 | 375061 | 1.9 | $0.93^{0.09}_{0.09}$ | – | 40.79 | $31.29^{0.05}_{0.06}$ | $30.95^{0.05}_{0.06}$ |
| 11 | 382482 | 1.7 | $0.99^{0.15}_{0.14}$ | – | 40.38 | $30.90^{0.09}_{0.08}$ | $30.54^{0.09}_{0.08}$ |
| 12 | 388558 | 2.9 | $0.90^{0.12}_{0.11}$ | – | 41.08 | $31.57^{0.07}_{0.07}$ | $31.24^{0.07}_{0.07}$ |
| 13 | 396371 | 2.0 | $0.77^{0.06}_{0.05}$ | – | 41.07 | $31.51^{0.04}_{0.04}$ | $31.22^{0.04}_{0.04}$ |
| 14 | 398977 | 1.6 | $1.02^{0.22}_{0.28}$ | $0.03^{0.29}_{0.34}$ | 40.12 | $30.63^{1.85}_{2.55}$ | $30.28^{1.85}_{2.55}$ |
| 15 | 406998 | 1.8 | $0.75^{0.24}_{0.21}$ | – | 40.19 | $30.61^{0.16}_{0.14}$ | $30.34^{0.16}_{0.14}$ |
| 16 | 410945 | 1.9 | $0.72^{0.09}_{0.09}$ | – | 40.80 | $31.21^{0.06}_{0.06}$ | $30.95^{0.06}_{0.06}$ |
| 17 | 411577 | 1.8 | $1.16^{0.17}_{0.16}$ | – | 40.39 | $30.97^{0.09}_{0.08}$ | $30.55^{0.09}_{0.08}$ |
| 18 | 415269 | 1.6 | $0.68^{0.23}_{0.19}$ | $0.05^{0.32}_{0.37}$ | 40.44 | $30.82^{1.86}_{2.56}$ | $30.58^{1.86}_{2.56}$ |
| 19 | 427827 | 2.4 | $0.66^{0.08}_{0.08}$ | – | 41.07 | $31.45^{0.06}_{0.06}$ | $31.21^{0.06}_{0.06}$ |
| 20 | 439437 | 3.4 | $0.74^{0.14}_{0.14}$ | – | 41.08 | $31.49^{0.10}_{0.10}$ | $31.22^{0.10}_{0.10}$ |
| 21 | 441820 | 3.0 | $0.71^{0.16}_{0.16}$ | – | 41.33 | $31.73^{0.12}_{0.12}$ | $31.47^{0.12}_{0.12}$ |
| 22 | 442090 | 1.6 | $0.46^{0.14}_{0.13}$ | – | 40.49 | $30.78^{0.15}_{0.14}$ | $30.61^{0.15}_{0.14}$ |
| 23 | 449963 | 2.6 | $0.36^{0.28}_{0.16}$ | – | 40.37 | $30.61^{0.34}_{0.21}$ | $30.48^{0.34}_{0.21}$ |
| 24 | 452505 | 1.9 | $0.51^{0.22}_{0.23}$ | – | 40.44 | $30.75^{0.22}_{0.22}$ | $30.56^{0.22}_{0.22}$ |
| 25 | 462193 | 1.9 | $0.88^{0.14}_{0.15}$ | – | 40.58 | $31.06^{0.09}_{0.10}$ | $30.74^{0.09}_{0.10}$ |
| 26 | 469122 | 3.2 | $0.89^{0.20}_{0.20}$ | – | 40.44 | $30.92^{0.12}_{0.12}$ | $30.59^{0.12}_{0.12}$ |
| 27 | 470051 | 1.9 | $0.48^{0.12}_{0.13}$ | – | 40.63 | $30.92^{0.12}_{0.12}$ | $30.75^{0.12}_{0.12}$ |
| 28 | 474662 | 2.0 | $0.78^{0.16}_{0.16}$ | – | 40.72 | $31.16^{0.11}_{0.11}$ | $30.87^{0.11}_{0.11}$ |
| 29 | 476581 | 3.2 | $0.78^{0.10}_{0.10}$ | – | 40.94 | $31.38^{0.06}_{0.07}$ | $31.09^{0.06}_{0.07}$ |
| 30 | 477095 | 1.8 | $0.75^{0.08}_{0.07}$ | – | 40.74 | $31.16^{0.05}_{0.05}$ | $30.89^{0.05}_{0.05}$ |
| 31 | 478096 | 1.8 | $1.04^{0.17}_{0.18}$ | – | 40.30 | $30.84^{0.1}_{0.1}$ | $30.46^{0.1}_{0.1}$ |
| 32 | 486997 | 1.8 | $0.79^{0.08}_{0.09}$ | – | 40.86 | $31.30^{0.05}_{0.06}$ | $31.01^{0.05}_{0.06}$ |
| 33 | 488383 | 1.8 | $0.43^{0.19}_{0.14}$ | – | 40.37 | $30.65^{0.20}_{0.15}$ | $30.49^{0.20}_{0.15}$ |
| 34 | 495704 | 2.1 | $0.67^{0.07}_{0.07}$ | – | 40.99 | $31.37^{0.06}_{0.06}$ | $31.13^{0.06}_{0.06}$ |
| 35 | 496707 | 2.0 | $0.52^{0.12}_{0.11}$ | – | 40.58 | $30.89^{0.11}_{0.10}$ | $30.70^{0.11}_{0.10}$ |
| 36 | 501111 | 2.4 | $0.62^{0.12}_{0.12}$ | – | 41.14 | $31.50^{0.10}_{0.1}$ | $31.27^{0.10}_{0.10}$ |
| 37 | 502532 | 1.9 | $0.81^{0.24}_{0.22}$ | – | 40.26 | $30.71^{0.15}_{0.14}$ | $30.41^{0.15}_{0.14}$ |
| 38 | 502934 | 1.7 | $1.01^{0.13}_{0.13}$ | – | 40.49 | $31.01^{0.07}_{0.07}$ | $30.65^{0.07}_{0.07}$ |
| 39 | 506989 | 1.6 | $0.95^{0.19}_{0.19}$ | – | 40.33 | $30.83^{0.11}_{0.11}$ | $30.49^{0.11}_{0.11}$ |
| 40 | 516049 | 1.6 | $1.13^{0.16}_{0.15}$ | – | 40.20 | $30.77^{0.09}_{0.09}$ | $30.36^{0.09}_{0.09}$ |
| 41 | 516610 | 1.6 | $0.58^{0.20}_{0.17}$ | – | 40.14 | $30.49^{0.16}_{0.14}$ | $30.28^{0.16}_{0.14}$ |
| 42 | 516810 | 1.6 | $0.71^{0.08}_{0.09}$ | – | 40.57 | $30.97^{0.06}_{0.06}$ | $30.72^{0.06}_{0.06}$ |

[1] Identifier from COSMOS Super-deblended Photometry Catalog (Jin et al. 2018)

[2] Photometric Redshift from COSMOS Catalogues (Weaver et al. 2022; Laigle et al. 2016)

| | | | | | | | |
|---|---|---|---|---|---|---|---|
| 43 | 525980 | 2.1 | 0.39 $_{0.15}^{0.18}$ | — | 40.44 | 30.70 $_{0.17}^{0.21}$ | 30.55 $_{0.17}^{0.21}$ |
| 44 | 526965 | 1.6 | 1.11 $_{0.12}^{0.13}$ | — | 40.28 | 30.84 $_{0.07}^{0.07}$ | 30.44 $_{0.07}^{0.07}$ |
| 45 | 529975 | 2.5 | 0.78 $_{0.17}^{0.19}$ | — | 40.72 | 31.15 $_{0.12}^{0.12}$ | 30.87 $_{0.12}^{0.12}$ |
| 46 | 533924 | 3.1 | 0.69 $_{0.13}^{0.11}$ | — | 41.09 | 31.49 $_{0.10}^{0.09}$ | 31.24 $_{0.10}^{0.09}$ |
| 47 | 540343 | 2.9 | 1.01 $_{0.13}^{0.12}$ | — | 40.36 | 30.89 $_{0.07}^{0.07}$ | 30.52 $_{0.07}^{0.07}$ |
| 48 | 547102 | 1.6 | 0.72 $_{0.14}^{0.14}$ | — | 40.31 | 30.71 $_{0.10}^{0.10}$ | 30.45 $_{0.10}^{0.10}$ |
| 49 | 556782 | 1.9 | 0.94 $_{0.10}^{0.11}$ | — | 40.43 | 30.93 $_{0.06}^{0.06}$ | 30.59 $_{0.06}^{0.06}$ |
| 50 | 559532 | 1.7 | 0.84 $_{0.17}^{0.16}$ | — | 40.28 | 30.74 $_{0.11}^{0.11}$ | 30.44 $_{0.11}^{0.11}$ |
| 51 | 561437 | 2.4 | 0.63 $_{0.14}^{0.12}$ | — | 40.95 | 31.32 $_{0.12}^{0.10}$ | 31.09 $_{0.10}^{0.10}$ |
| 52 | 564732 | 1.8 | 0.69 $_{0.11}^{0.12}$ | — | 40.55 | 30.95 $_{0.08}^{0.08}$ | 30.70 $_{0.08}^{0.08}$ |
| 53 | 565719 | 1.9 | 0.87 $_{0.13}^{0.12}$ | — | 40.62 | 31.09 $_{0.08}^{0.08}$ | 30.77 $_{0.08}^{0.08}$ |
| 54 | 565792 | 1.9 | 0.98 $_{0.08}^{0.08}$ | — | 40.81 | 31.33 $_{0.05}^{0.05}$ | 30.97 $_{0.05}^{0.05}$ |
| 55 | 569615 | 2.8 | 0.73 $_{0.09}^{0.11}$ | — | 40.91 | 31.32 $_{0.07}^{0.08}$ | 31.06 $_{0.08}^{0.08}$ |
| 56 | 575519 | 1.9 | 0.58 $_{0.19}^{0.16}$ | — | 40.38 | 30.72 $_{0.16}^{0.14}$ | 30.51 $_{0.16}^{0.14}$ |
| 57 | 578926 | 2.4 | 0.81 $_{0.08}^{0.08}$ | — | 41.02 | 31.46 $_{0.05}^{0.05}$ | 31.17 $_{0.05}^{0.05}$ |
| 58 | 579041 | 1.8 | 0.71 $_{0.13}^{0.11}$ | — | 40.57 | 30.97 $_{0.09}^{0.08}$ | 30.71 $_{0.09}^{0.08}$ |
| 59 | 580926 | 1.6 | 0.87 $_{0.18}^{0.18}$ | — | 40.15 | 30.62 $_{0.11}^{0.12}$ | 30.31 $_{0.11}^{0.12}$ |
| 60 | 584422 | 2.8 | 0.64 $_{0.09}^{0.10}$ | — | 41.24 | 31.61 $_{0.07}^{0.07}$ | 31.38 $_{0.07}^{0.07}$ |
| 61 | 590368 | 2.4 | 0.61 $_{0.17}^{0.18}$ | — | 40.86 | 31.22 $_{0.15}^{0.16}$ | 31.00 $_{0.15}^{0.16}$ |
| 62 | 595018 | 1.7 | 0.65 $_{0.12}^{0.11}$ | — | 40.49 | 30.87 $_{0.10}^{0.09}$ | 30.63 $_{0.10}^{0.09}$ |
| 63 | 595222 | 1.6 | 0.64 $_{0.10}^{0.11}$ | — | 40.35 | 30.72 $_{0.08}^{0.08}$ | 30.49 $_{0.08}^{0.08}$ |
| 64 | 597580 | 2.0 | 0.81 $_{0.21}^{0.17}$ | 0.10 $_{0.38}^{0.25}$ | 40.76 | 31.18 $_{2.20}^{1.44}$ | 30.90 $_{2.20}^{1.44}$ |
| 65 | 612581 | 1.8 | 0.74 $_{0.14}^{0.12}$ | — | 40.48 | 30.90 $_{0.10}^{0.09}$ | 30.63 $_{0.09}^{0.09}$ |
| 66 | 616280 | 3.0 | 0.52 $_{0.06}^{0.06}$ | — | 41.46 | 31.77 $_{0.06}^{0.06}$ | 31.58 $_{0.06}^{0.06}$ |
| 67 | 623091 | 2.2 | 0.79 $_{0.11}^{0.10}$ | — | 40.85 | 31.29 $_{0.07}^{0.07}$ | 31.00 $_{0.07}^{0.07}$ |
| 68 | 625223 | 1.8 | 0.35 $_{0.10}^{0.11}$ | — | 40.45 | 30.68 $_{0.13}^{0.14}$ | 30.55 $_{0.13}^{0.14}$ |
| 69 | 627356 | 2.2 | 0.67 $_{0.06}^{0.06}$ | — | 41.33 | 31.72 $_{0.04}^{0.04}$ | 31.47 $_{0.04}^{0.04}$ |
| 70 | 628033 | 1.8 | 0.81 $_{0.10}^{0.10}$ | — | 40.81 | 31.26 $_{0.07}^{0.07}$ | 30.96 $_{0.07}^{0.07}$ |
| 71 | 631682 | 3.3 | 0.65 $_{0.10}^{0.10}$ | — | 41.14 | 31.51 $_{0.08}^{0.08}$ | 31.28 $_{0.08}^{0.08}$ |
| 72 | 632758 | 1.8 | 1.02 $_{0.07}^{0.07}$ | — | 40.69 | 31.22 $_{0.04}^{0.04}$ | 30.85 $_{0.04}^{0.04}$ |
| 73 | 634466 | 3.3 | 0.79 $_{0.20}^{0.19}$ | — | 41.01 | 31.44 $_{0.14}^{0.13}$ | 31.16 $_{0.14}^{0.13}$ |
| 74 | 634999 | 1.8 | 1.11 $_{0.10}^{0.10}$ | — | 40.38 | 30.94 $_{0.05}^{0.06}$ | 30.54 $_{0.05}^{0.06}$ |
| 75 | 635353 | 2.6 | 0.98 $_{0.14}^{0.15}$ | — | 40.82 | 31.34 $_{0.09}^{0.09}$ | 30.98 $_{0.09}^{0.09}$ |
| 76 | 635836 | 1.7 | 0.82 $_{0.09}^{0.10}$ | — | 40.55 | 31.00 $_{0.06}^{0.07}$ | 30.71 $_{0.06}^{0.07}$ |
| 77 | 638136 | 1.5 | 0.93 $_{0.28}^{0.28}$ | 0.06 $_{0.44}^{0.27}$ | 40.22 | 30.70 $_{3.23}^{2.42}$ | 30.37 $_{3.23}^{2.42}$ |
| 78 | 638642 | 2.0 | 0.62 $_{0.13}^{0.13}$ | — | 40.53 | 30.90 $_{0.11}^{0.10}$ | 30.67 $_{0.11}^{0.10}$ |
| 79 | 644839 | 1.6 | 0.95 $_{0.15}^{0.14}$ | — | 40.40 | 30.90 $_{0.09}^{0.09}$ | 30.56 $_{0.09}^{0.09}$ |
| 80 | 645724 | 2.3 | 0.82 $_{0.08}^{0.08}$ | — | 41.20 | 31.65 $_{0.05}^{0.05}$ | 31.35 $_{0.05}^{0.05}$ |
| 81 | 647970 | 1.9 | 1.08 $_{0.24}^{0.21}$ | — | 40.38 | 30.93 $_{0.14}^{0.13}$ | 30.54 $_{0.14}^{0.13}$ |
| 82 | 651173 | 1.8 | 0.84 $_{0.08}^{0.07}$ | — | 40.84 | 31.30 $_{0.05}^{0.05}$ | 30.99 $_{0.05}^{0.05}$ |
| 83 | 651584 | 2.3 | 0.59 $_{0.10}^{0.10}$ | — | 40.97 | 31.32 $_{0.09}^{0.09}$ | 31.10 $_{0.09}^{0.09}$ |
| 84 | 657953 | 1.5 | 0.79 $_{0.15}^{0.18}$ | 0.17 $_{0.31}^{0.22}$ | 40.60 | 30.98 $_{0.54}^{0.45}$ | 30.73 $_{0.54}^{0.45}$ |
| 85 | 662169 | 2.2 | 1.22 $_{0.12}^{0.10}$ | — | 40.64 | 31.25 $_{0.08}^{0.07}$ | 30.81 $_{0.08}^{0.07}$ |
| 86 | 662536 | 2.6 | 0.74 $_{0.11}^{0.10}$ | — | 40.98 | 31.40 $_{0.08}^{0.08}$ | 31.13 $_{0.08}^{0.08}$ |
| 87 | 678384 | 1.8 | 0.84 $_{0.15}^{0.13}$ | — | 40.52 | 30.98 $_{0.10}^{0.09}$ | 30.67 $_{0.10}^{0.09}$ |
| 88 | 680956 | 2.7 | 0.88 $_{0.11}^{0.10}$ | — | 41.04 | 31.51 $_{0.07}^{0.07}$ | 31.19 $_{0.07}^{0.07}$ |
| 89 | 690848 | 2.0 | 0.76 $_{0.16}^{0.17}$ | — | 40.60 | 31.02 $_{0.12}^{0.12}$ | 30.75 $_{0.12}^{0.12}$ |
| 90 | 692002 | 2.9 | 0.67 $_{0.10}^{0.10}$ | — | 41.27 | 31.65 $_{0.08}^{0.08}$ | 31.41 $_{0.08}^{0.08}$ |

| 91 | 696690 | 2.7 | $1.20_{0.14}^{0.13}$ | − | 40.59 | $31.18_{0.07}^{0.07}$ | $30.75_{0.07}^{0.07}$ |
|---|---|---|---|---|---|---|---|
| 92 | 705179 | 1.7 | $0.84_{0.09}^{0.10}$ | − | 40.59 | $31.05_{0.06}^{0.06}$ | $30.74_{0.06}^{0.06}$ |
| 93 | 705616 | 1.9 | $0.85_{0.13}^{0.13}$ | − | 40.64 | $31.10_{0.08}^{0.08}$ | $30.79_{0.08}^{0.08}$ |
| 94 | 707850 | 3.2 | $0.31_{0.12}^{0.16}$ | − | 40.82 | $31.03_{0.17}^{0.23}$ | $30.92_{0.17}^{0.23}$ |
| 95 | 708131 | 2.0 | $0.69_{0.06}^{0.07}$ | − | 41.06 | $31.45_{0.05}^{0.05}$ | $31.20_{0.05}^{0.05}$ |
| 96 | 710739 | 1.9 | $1.09_{0.33}^{0.24}$ | $0.06_{0.27}^{0.20}$ | 40.45 | $30.99_{2.89}^{2.07}$ | $30.60_{2.89}^{2.07}$ |
| 97 | 717183 | 1.9 | $1.04_{0.15}^{0.12}$ | − | 40.54 | $31.08_{0.08}^{0.08}$ | $30.70_{0.08}^{0.08}$ |
| 98 | 718344 | 2.9 | $0.72_{0.16}^{0.16}$ | − | 40.84 | $31.24_{0.11}^{0.11}$ | $30.98_{0.11}^{0.11}$ |
| 99 | 727184 | 2.0 | $0.79_{0.15}^{0.15}$ | − | 40.56 | $31.00_{0.10}^{0.10}$ | $30.71_{0.10}^{0.10}$ |
| 100 | 729926 | 1.8 | $0.86_{0.17}^{0.18}$ | − | 40.37 | $30.84_{0.11}^{0.11}$ | $30.53_{0.11}^{0.11}$ |
| 101 | 730319 | 2.9 | $0.50_{0.07}^{0.08}$ | − | 41.17 | $31.48_{0.07}^{0.08}$ | $31.30_{0.07}^{0.08}$ |
| 102 | 732468 | 2.7 | $0.57_{0.22}^{0.23}$ | − | 40.69 | $31.03_{0.20}^{0.20}$ | $30.83_{0.20}^{0.20}$ |
| 103 | 737676 | 1.8 | $0.85_{0.13}^{0.12}$ | − | 40.62 | $31.08_{0.09}^{0.08}$ | $30.78_{0.09}^{0.08}$ |
| 104 | 748925 | 3.2 | $1.06_{0.18}^{0.17}$ | − | 40.37 | $30.91_{0.10}^{0.09}$ | $30.53_{0.10}^{0.09}$ |
| 105 | 749578 | 1.7 | $0.95_{0.07}^{0.07}$ | − | 40.82 | $31.32_{0.04}^{0.04}$ | $30.98_{0.04}^{0.04}$ |
| 106 | 751335 | 1.6 | $0.73_{0.10}^{0.08}$ | − | 40.73 | $31.15_{0.07}^{0.07}$ | $30.88_{0.07}^{0.07}$ |
| 107 | 757799 | 1.6 | $0.69_{0.13}^{0.12}$ | − | 40.37 | $30.77_{0.09}^{0.09}$ | $30.52_{0.09}^{0.09}$ |
| 108 | 762527 | 2.1 | $0.80_{0.11}^{0.13}$ | − | 40.36 | $30.80_{0.08}^{0.09}$ | $30.51_{0.08}^{0.09}$ |
| 109 | 770288 | 2.1 | $0.40_{0.12}^{0.11}$ | − | 40.62 | $30.87_{0.13}^{0.13}$ | $30.73_{0.13}^{0.13}$ |
| 110 | 772906 | 2.0 | $0.36_{0.09}^{0.10}$ | − | 40.81 | $31.05_{0.11}^{0.12}$ | $30.92_{0.11}^{0.12}$ |
| 111 | 791065 | 2.0 | $0.26_{0.13}^{0.18}$ | $0.14_{0.61}^{0.51}$ | 41.03 | $31.20_{1.94}^{1.38}$ | $31.12_{1.94}^{1.38}$ |
| 112 | 799864 | 2.4 | $0.83_{0.10}^{0.09}$ | − | 40.84 | $31.30_{0.06}^{0.06}$ | $31.00_{0.06}^{0.06}$ |
| 113 | 801243 | 1.8 | $0.73_{0.10}^{0.11}$ | − | 40.58 | $30.99_{0.07}^{0.07}$ | $30.73_{0.07}^{0.07}$ |
| 114 | 803543 | 2.6 | $0.86_{0.17}^{0.15}$ | − | 40.81 | $31.28_{0.11}^{0.11}$ | $30.97_{0.11}^{0.11}$ |
| 115 | 808735 | 1.5 | $0.22_{0.10}^{0.30}$ | − | 40.13 | $30.30_{0.21}^{0.61}$ | $30.22_{0.21}^{0.61}$ |
| 116 | 819326 | 2.6 | $0.62_{0.14}^{0.13}$ | − | 40.93 | $31.30_{0.12}^{0.11}$ | $31.07_{0.12}^{0.11}$ |
| 117 | 827300 | 1.5 | $0.78_{0.09}^{0.09}$ | − | 40.47 | $30.90_{0.06}^{0.06}$ | $30.62_{0.06}^{0.06}$ |
| 118 | 837455 | 1.8 | $0.62_{0.15}^{0.13}$ | − | 40.59 | $30.95_{0.13}^{0.11}$ | $30.73_{0.13}^{0.11}$ |
| 119 | 838572 | 2.7 | $0.42_{0.19}^{0.22}$ | $0.10_{0.59}^{0.45}$ | 40.83 | $31.08_{2.13}^{1.45}$ | $30.94_{2.13}^{1.45}$ |
| 120 | 842239 | 1.7 | $0.92_{0.14}^{0.15}$ | − | 40.38 | $30.87_{0.09}^{0.10}$ | $30.54_{0.09}^{0.10}$ |
| 121 | 842950 | 2.5 | $0.80_{0.14}^{0.16}$ | − | 40.76 | $31.20_{0.10}^{0.11}$ | $30.92_{0.10}^{0.11}$ |
| 122 | 849028 | 2.9 | $0.62_{0.08}^{0.07}$ | − | 41.30 | $31.66_{0.07}^{0.06}$ | $31.44_{0.07}^{0.06}$ |
| 123 | 851363 | 2.0 | $0.79_{0.05}^{0.05}$ | − | 41.06 | $31.50_{0.03}^{0.03}$ | $31.21_{0.03}^{0.03}$ |
| 124 | 852900 | 2.4 | $0.36_{0.15}^{0.28}$ | − | 40.43 | $30.67_{0.20}^{0.35}$ | $30.54_{0.20}^{0.35}$ |
| 125 | 863440 | 1.6 | $1.25_{0.30}^{0.15}$ | $0.07_{0.18}^{0.16}$ | 40.28 | $30.89_{4.07}^{3.09}$ | $30.44_{4.07}^{3.09}$ |
| 126 | 871000 | 1.7 | $1.31_{0.11}^{0.11}$ | − | 40.69 | $31.33_{0.06}^{0.06}$ | $30.85_{0.06}^{0.06}$ |
| 127 | 871582 | 1.5 | $1.15_{0.16}^{0.15}$ | − | 40.32 | $30.89_{0.09}^{0.09}$ | $30.48_{0.09}^{0.09}$ |
| 128 | 896241 | 1.8 | $0.96_{0.16}^{0.14}$ | − | 40.54 | $31.05_{0.10}^{0.09}$ | $30.70_{0.10}^{0.09}$ |
| 129 | 902320 | 1.7 | $0.81_{0.06}^{0.06}$ | − | 41.09 | $31.53_{0.04}^{0.04}$ | $31.24_{0.04}^{0.04}$ |
| 130 | 10033174 | 2.8 | $0.74_{0.12}^{0.13}$ | − | 40.99 | $31.41_{0.09}^{0.09}$ | $31.14_{0.09}^{0.09}$ |
| 131 | 10035127 | 2.0 | $0.57_{0.08}^{0.08}$ | − | 40.93 | $31.27_{0.07}^{0.07}$ | $31.06_{0.07}^{0.07}$ |
| 132 | 10040313 | 1.8 | $0.84_{0.09}^{0.09}$ | − | 40.66 | $31.12_{0.06}^{0.05}$ | $30.82_{0.06}^{0.05}$ |
| 133 | 10070338 | 1.7 | $0.48_{0.09}^{0.09}$ | − | 40.67 | $30.97_{0.08}^{0.08}$ | $30.80_{0.08}^{0.08}$ |
| 134 | 10078195 | 1.6 | $0.66_{0.08}^{0.08}$ | − | 40.67 | $31.05_{0.06}^{0.06}$ | $30.81_{0.06}^{0.06}$ |
| 135 | 10080288 | 2.3 | $0.77_{0.11}^{0.12}$ | − | 40.86 | $31.29_{0.08}^{0.08}$ | $31.01_{0.08}^{0.08}$ |
| 136 | 10085717 | 1.7 | $0.89_{0.27}^{0.28}$ | $0.14_{0.36}^{0.24}$ | 40.44 | $30.87_{0.89}^{0.60}$ | $30.58_{0.89}^{0.6}$ |
| 137 | 10090559 | 1.5 | $0.75_{0.12}^{0.14}$ | − | 40.26 | $30.69_{0.09}^{0.10}$ | $30.41_{0.09}^{0.10}$ |
| 138 | 10090785 | 2.0 | $0.69_{0.10}^{0.10}$ | − | 40.76 | $31.15_{0.08}^{0.08}$ | $30.90_{0.08}^{0.08}$ |
| 139 | 10092310 | 1.8 | $0.80^{0.08}_{0.09}$ | — | 40.74 | $31.19^{0.06}_{0.06}$ | $30.90^{0.06}_{0.06}$ |
| 140 | 10093049 | 1.6 | $0.85^{0.09}_{0.10}$ | — | 40.43 | $30.89^{0.06}_{0.07}$ | $30.59^{0.06}_{0.07}$ |
| 141 | 10093193 | 2.3 | $0.71^{0.08}_{0.10}$ | — | 41.17 | $31.57^{0.06}_{0.07}$ | $31.31^{0.06}_{0.07}$ |
| 142 | 10094402 | 1.8 | $0.93^{0.09}_{0.11}$ | — | 40.76 | $31.26^{0.06}_{0.07}$ | $30.92^{0.06}_{0.07}$ |
| 143 | 10094777 | 1.6 | $0.91^{0.08}_{0.08}$ | — | 40.65 | $31.14^{0.05}_{0.05}$ | $30.80^{0.05}_{0.05}$ |
| 144 | 10095574 | 1.8 | $0.71^{0.09}_{0.09}$ | — | 40.70 | $31.11^{0.07}_{0.06}$ | $30.85^{0.07}_{0.06}$ |
| 145 | 10103325 | 2.3 | $0.70^{0.06}_{0.06}$ | — | 41.26 | $31.66^{0.04}_{0.05}$ | $31.41^{0.04}_{0.05}$ |
| 146 | 10123324 | 2.3 | $0.62^{0.08}_{0.08}$ | — | 40.92 | $31.28^{0.06}_{0.07}$ | $31.06^{0.06}_{0.07}$ |
| 147 | 10124598 | 2.1 | $0.63^{0.09}_{0.08}$ | — | 40.89 | $31.25^{0.07}_{0.07}$ | $31.03^{0.07}_{0.07}$ |
| 148 | 10134094 | 2.0 | $0.65^{0.09}_{0.08}$ | — | 40.80 | $31.17^{0.07}_{0.07}$ | $30.94^{0.07}_{0.07}$ |
| 149 | 10135986 | 2.0 | $1.00^{0.14}_{0.14}$ | — | 40.53 | $31.06^{0.08}_{0.08}$ | $30.69^{0.08}_{0.08}$ |
| 150 | 10137050 | 2.0 | $0.32^{0.13}_{0.11}$ | — | 40.62 | $30.84^{0.18}_{0.15}$ | $30.72^{0.18}_{0.15}$ |
| 151 | 10138523 | 3.0 | $0.50^{0.16}_{0.13}$ | — | 40.88 | $31.19^{0.15}_{0.13}$ | $31.00^{0.15}_{0.13}$ |
| 152 | 10139765 | 1.7 | $0.71^{0.09}_{0.08}$ | — | 40.72 | $31.12^{0.06}_{0.06}$ | $30.87^{0.06}_{0.06}$ |
| 153 | 10149158 | 2.5 | $0.76^{0.06}_{0.06}$ | — | 41.35 | $31.77^{0.05}_{0.05}$ | $31.50^{0.05}_{0.05}$ |
| 154 | 10157709 | 1.9 | $0.93^{0.10}_{0.10}$ | — | 40.72 | $31.22^{0.06}_{0.06}$ | $30.88^{0.06}_{0.06}$ |
| 155 | 10176688 | 1.5 | $0.70^{0.07}_{0.07}$ | — | 40.61 | $31.01^{0.05}_{0.05}$ | $30.76^{0.05}_{0.05}$ |
| 156 | 10205913 | 2.2 | $0.53^{0.12}_{0.13}$ | — | 40.72 | $31.04^{0.11}_{0.12}$ | $30.85^{0.11}_{0.12}$ |
| 157 | 10209718 | 1.5 | $0.79^{0.08}_{0.10}$ | — | 40.50 | $30.94^{0.06}_{0.07}$ | $30.65^{0.06}_{0.07}$ |
| 158 | 10211581 | 3.2 | $0.66^{0.25}_{0.23}$ | — | 40.79 | $31.17^{0.20}_{0.18}$ | $30.93^{0.20}_{0.18}$ |
| 159 | 10219023 | 2.6 | $0.64^{0.07}_{0.06}$ | — | 41.23 | $31.60^{0.05}_{0.05}$ | $31.37^{0.05}_{0.05}$ |
| 160 | 10219338 | 1.8 | $0.92^{0.22}_{0.22}$ | — | 40.27 | $30.76^{0.13}_{0.14}$ | $30.43^{0.13}_{0.14}$ |

## B. THERMAL FRACTIONS USING DE-REDDEND H$\alpha$ EMISSION

This study combines data from FMOS-COSMOS line emission (Silverman et al. 2015; Kashino et al. 2019) with the MIGHTEE continuum survey. The FMOS-COSMOS is a large near-infrared (NIR) spectroscopic survey of star-forming galaxies and AGNs in the COSMOS field with the Fiber-Multi Object Spectrograph (FMOS) mounted on the Subaru Telescope. The FMOS H band spectral window covers the spectral range of 1.6-1.8 $\mu$m to detect the rest-frame H$\alpha$ and [NII]$\lambda$6584 lines and the 1.11-1.35 $\mu$m range to detect the rest-frame H$\beta$ and [OIII]$\lambda$5008 lines. The analysis focuses on galaxies in the redshift range $1.5 < z < 2$, excluding sources with null or negative H$\alpha$ or H$\beta$ fluxes. Cross-matching the FMOS-COMOS with our sample using $TOPCAT$ with a matching criterion of < 1" separation, we found six common SFGs (Table A3). The aperture corrected observed fluxes of the pure H$_\alpha$ and H$_\beta$ lines (S$_{H\alpha}$ and S$_{H\beta}$) were used to derive color excess $E(B-V)$ following Osterbrock & Ferland (2006),

$$E(B-V) = 1.97 \log_{10}\left(\frac{S_{H\alpha}/S_{H\beta}}{2.86}\right).$$

The extinction values of $A_{H_\alpha}$ is then obtained assuming a Milky-Way like reddening (Calzetti et al. 2000),

$$A_{H\alpha} = (3.33 \pm 0.80) \times E(B-V).$$

The de-reddened H$\alpha$ flux, $S_{H\alpha} = 10^{A_{H\alpha}} (S_{H\alpha})_{\text{obs}}$, is then converted to that of the free-free radio emission, $S_\nu^{\text{th}}$, at frequency $\nu$ via

$$\left(\frac{S_\nu^{\text{th}}}{\text{erg s}^{-1}\,\text{cm}^{-2}\text{Hz}^{-1}}\right) = 1.14 \times 10^{-14} \left(\frac{T_e}{10^4\,\text{K}}\right)^{0.34} \left(\frac{\nu}{\text{GHz}}\right)^{-0.1} \frac{S_{H\alpha}}{\text{erg s}^{-1}\,\text{cm}^{-2}}.$$

We obtained the thermal flux at 1.3 GHz observed frame, $S_{1.3}^{\text{th}}$, for an electron temperature of $T_e = 10^4$ K. The resulting thermal fractions, $f_{\text{th}}^{H\alpha} = S_{1.3}^{\text{th}}/S_{1.3}$ of these SFGs are listed in Table A3. Sources indicated by $^\star$ in the table have $S_{H\alpha}/S_{H\beta} < 2.86$ and, thus, no $A_{H\alpha}$ can be determined for. As a result, $f_{\text{th}}^{H\alpha}$ of these sources is a lower limit to their thermal fraction. In general, $f_{\text{th}}^{H\alpha}$ agrees with that obtained using the radio SED method, $f_{\text{th}}^{\text{SED}}$, taking into account uncertainties.

Table A3: Thermal fractions of SFGs at 1.3 GHz observed frame obtained using the de-reddened H$\alpha$ ($f_{\text{th}}^{H_\alpha}$) and the SED ($f_{\text{th}}^{\text{SED}}$) methods.

| ID | Redshift | $A_{H_\alpha}$ | $f_{\text{th}}^{H_\alpha}$ | $f_{\text{th}}^{\text{SED}}$ |
|---|---|---|---|---|
|  |  |  | % | % |
| 350085$^\star$ | 1.6 | – | 0.01$\pm$0.00 | – |
| 398977$^\star$ | 1.5 | – | 0.01$\pm$0.01 | 0.03$_{0.34}^{0.29}$ |
| 415269 | 1.6 | 2.05 | 0.06$\pm$0.03 | 0.05$_{0.37}^{0.32}$ |
| 757799 | 1.6 | 0.24 | 0.01$\pm$0.00 | – |
| 863440$^\star$ | 1.6 | – | 0.01$\pm$0.00 | 0.07$_{0.18}^{0.16}$ |
| 902320 | 1.7 | 2.9 | 0.02$\pm$0.01 | – |

## C. TEST OF SAMPLE SELECTION BIAS

In this section, we test whether selecting galaxies with only robust SEDs, i.e., those fitted with all 5 data points with SNR> 1 causes a bias on the results in 3 ways.

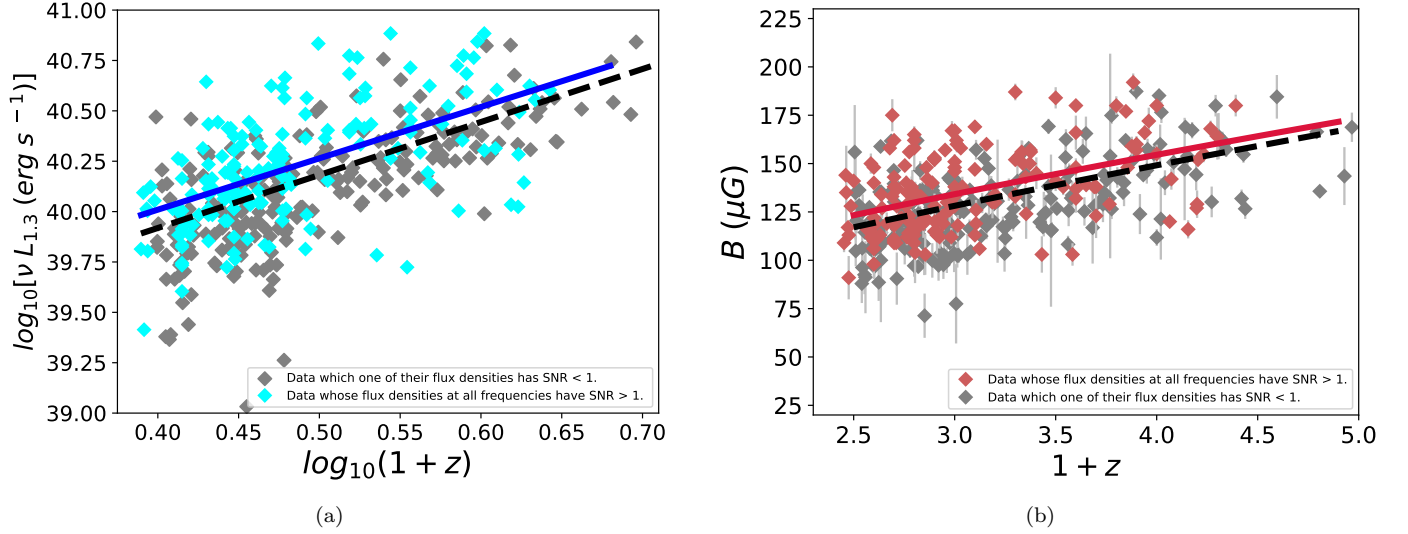

Figure 13: Cosmic evolution of the radio luminosity at $\nu\,L_{1.3}$ (a) and the magnetic field strength (b) of the MIGHTEE-COSMOS SFGs at $1.5 < z < 4$. Also shown are the linear regression fits (left panel) and the best curve fits (right panel) to the data for galaxies whose flux densities at all frequencies have a signal-to-noise ratio larger than one (SNR > 1, solid lines) and to all data (dashed lines).

### C.1. Test 1:

We compare the evolution of $\nu\,L_{1.3}$ for the case of only 160 galaxies with robust SEDs (blue points in Fig. 13) with the case in which also other galaxies having at least one flux density $< 1\sigma$ (gray points in Fig. 13) are included. A linear regression fit to the first case leads to the following relation:

$$\log_{10}(\nu\,L_{1.3}) = (2.9 \pm 0.2) \log_{10}(1+z) + (38.81 \pm 0.12), \tag{C1}$$

while in the second case,

$$\log_{10}(\nu\,L_{1.3}) = (2.8 \pm 0.1) \log_{10}(1+z) + (38.79 \pm 0.07). \tag{C2}$$

The two fitted relations agree with each other within the errors. Figure 13-right shows the evolution of the magnetic field strength for the two cases. In the first case, we find

$$B = (47 \pm 8) \times (1+z)^{(0.72 \pm 0.10)}, \tag{C3}$$

and in the second case,

$$B = (40 \pm 4) \times (1+z)^{(0.79 \pm 0.05)}, \tag{C4}$$

showing again an agreement.

### C.2. Test 2:

To ensure that we are less biased to the shape of the spectrum, we increase the flux limit of the lowest frequency from $1\sigma$ to $2\sigma$ and $3\sigma$ for selecting the sample. By increasing the flux limit we should still be able to detect the steepest sources ($\alpha_{\rm nt} > 0.8$) which is the case (see Fig. 14).

### C.3. Test 3:

The histogram distributions of the non-thermal spectral index for different flux density ranges at 1.3 GHz have been shown in Fig. 15. These distributions almost cover all possible non-thermal spectral index ranges. $F_{1.3}$ is flux density at 1.3 GHz.

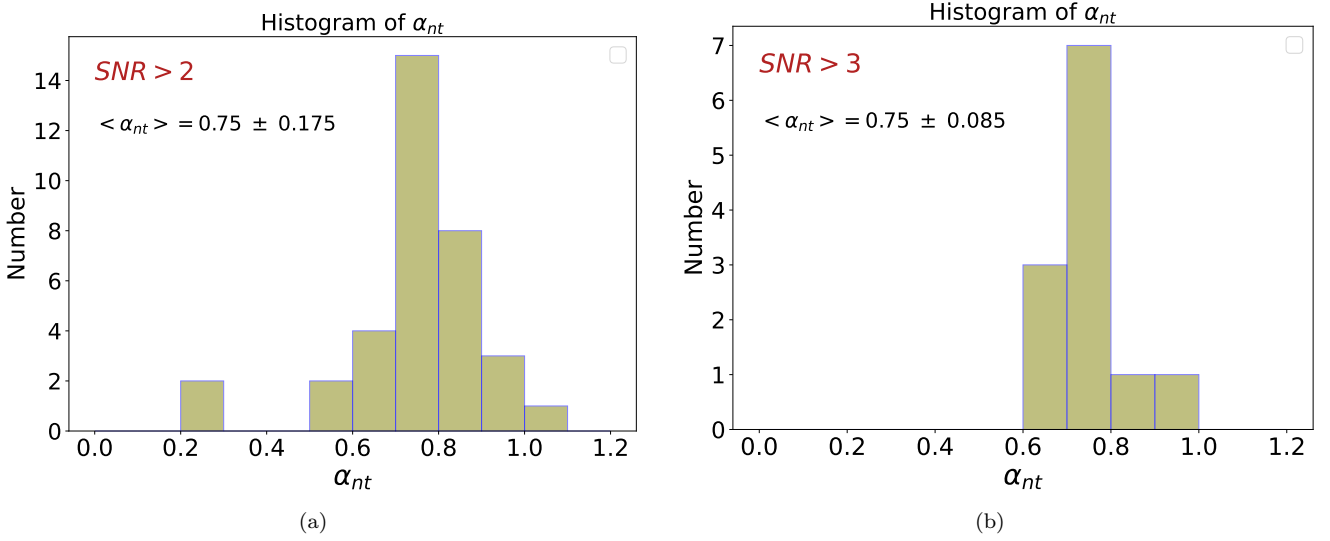

Figure 14: Histrogram of the nonthermal spectral index $\alpha_{\rm nt}$ for two different signal-to-noise ratio limits SNR$> 2$ (left) and SNR$> 3$ (right) at lowest frequency.

## D. EQUIPARTITION MAGNETIC FIELD STRENGTH

Assuming the equipartition between the magnetic field energy density and that of cosmic rays ($E_B = E_{CR} = B_{eq}^2/8\pi$), the strength of the galactic magnetic field B in Gauss is given by

$$B = \left[ \frac{4\pi(2\alpha_{nt}+1)K'I_{nt}E_p^{(1-2\alpha_{nt})}(\frac{\nu}{2c_1})^{\alpha_{nt}}}{(2\alpha_{nt}-1)\,c_2\,l\,c_3} \right]^{\frac{1}{\alpha_{nt}+3}}, \tag{D5}$$

where K is the ratio between the number densities of cosmic-ray protons and electrons and $K' = K + 1$ with K the ratio between the number densities of cosmic-ray protons and electrons. $I_{nt}$ is the nonthermal intensity in erg s$^{-1}$ cm$^{-2}$ Hz$^{-1}$ sr$^{-1}$, $\alpha_{nt}$ is the mean synchrotron spectral index, $l$ is pathlength through the nonthermal synchrotron emitting medium in cm ($l \simeq 1\ kpc/\cos i$), K $\simeq$ 100 (Beck & Krause 2005), and $E_p = 938.26$ MeV $= 1.50 \times 10^{-3}$ erg is the proton rest energy. The constants $c_1$, $c_2$ and $c_4$ are given as:

$$c_1 = \frac{3e}{4\pi m_e^3 c^5} = 6.26428 \times 10^{18}\ {\rm erg}^{-2}\ {\rm s}^{-1}\ {\rm G}^{-1},$$

$$c_2(\alpha_{nt}) = \frac{1}{4}c_4 \frac{(\alpha_{nt}+\frac{5}{3})}{(\alpha_{nt}+1)})\Gamma[\frac{3\alpha_{nt}+1}{6}] \times \Gamma[\frac{3\alpha_{nt}+5}{6}],$$

$$c_4 = \sqrt{3}\frac{e^3}{4\pi m_e c^2} = 1.86558 \times 10^{-23}\ {\rm erg\ sr}^{-1}\ {\rm G}^{-1} \tag{D6}$$

in which $\Gamma$ is the mathematical gamma function. For a uniform field with a constant inclination $i$ with respect to the sky plane ($i = 0°$ is the face-on view), $c_3 = [\cos(i)]^{(1+\alpha_{\rm nt})}$. For dominant synchrotron cooling of the cosmic ray electrons, Eq. (D5) is reduced to $B = C\left(\frac{I_{\rm nt}}{\cos(i)}\right)^{1/4}$ in which C is constant.

The specific intensity of a galaxy $I_{\rm nt}$ is related to its flux density $S_{nt}$ as $I_{\rm nt} \sim S_{\rm nt}/R_e^2$, with $R_e$ its effective radius. The ratio of the equipartition magnetic field strengths of two galaxies can then be estimated if the ratio of their flux densities (after correcting for their different distances or equivalently their luminosity ratio) per ratio of their areas is known. Taking into account the size-mass relation, $R_e \propto M_\star^{0.2}$ (e.g., van der Wel et al. 2014; Du et al. 2024), we can

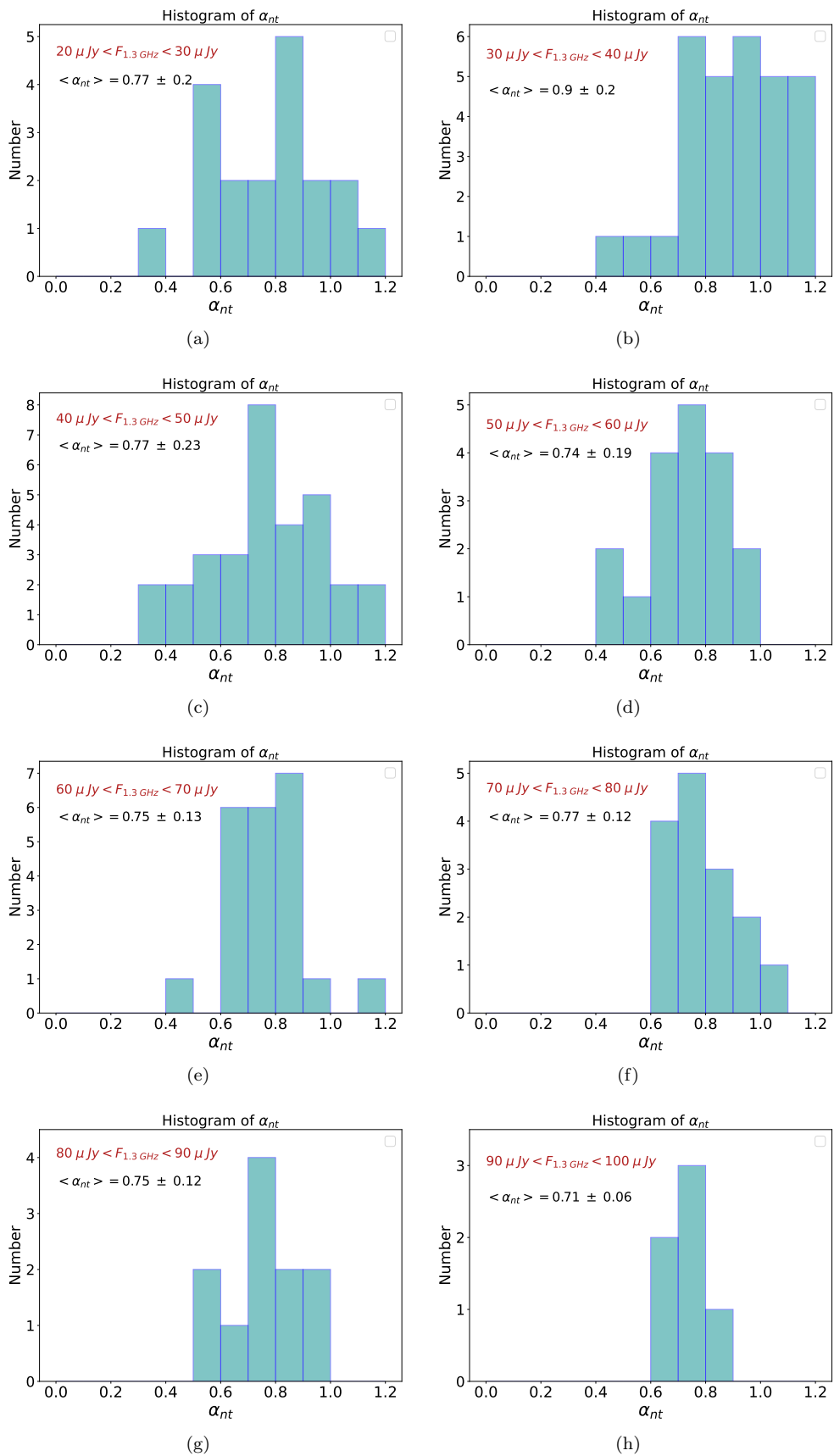

Figure 15: The non-thermal spectral index distribution for different flux density ranges at 1.3 GHz.

obtain the following useful expression:

$$B = B_0 \left(\frac{\cos(i)}{\cos(i_0)}\right)^{-1/4} \left(\frac{L_{nt}}{L_{nt0}}\right)^{1/4} \left(\frac{M_\star}{M_{\star 0}}\right)^{-0.1}, \tag{D7}$$

with index '0' referring to a reference galaxy (at z=0 in this study, see Sect. 5) with known magnetic field strength $B_0$.